\documentclass[aps,pre,onecolumn,superscriptaddress,showpacs]{revtex4}

\usepackage{amssymb}
\usepackage{epsfig}
\usepackage{graphicx}
\usepackage{float}
\usepackage{amsmath}
\usepackage{times}
\usepackage{multirow}
\usepackage{caption}
\usepackage{graphicx,epstopdf}
\usepackage{xcolor}

\begin{document}

\title{Theoretical results on the topological properties of the limited
  penetrable horizontal visibility graph family}
\author{Minggang Wang}\email{magic821204@sina.com, magic82@bu.edu}
\affiliation{School of Mathematical Science, Nanjing Normal University,
  Nanjing 210042, Jiangsu, China}
\affiliation{Department of Mathematics, Nanjing Normal University
  Taizhou College, Taizhou 225300, Jiangsu, China}
\affiliation{Center for Polymer Studies and Department of Physics,
  Boston University, Boston, MA 02215, USA}

\author{Andr\'{e} L.M.Vilela}
\affiliation{Center for Polymer Studies and Department of Physics,
  Boston University, Boston, MA 02215, USA}
\affiliation{Universidade de Pernambuco, 50720-001, Recife-PE, Brazil}

\author{Ruijin Du}
\affiliation{Center for Polymer Studies and Department of Physics,
  Boston University, Boston, MA 02215, USA}
\affiliation{Energy Development and Environmental Protection Strategy
  Research Center, Jiangsu University, Zhenjiang, 212013 Jiangsu, China}

\author{Longfeng Zhao}
\affiliation{Center for Polymer Studies and Department of Physics,
  Boston University, Boston, MA 02215, USA}

\author{Gaogao Dong}
\affiliation{Center for Polymer Studies and Department of Physics,
  Boston University, Boston, MA 02215, USA}
\affiliation{Energy Development and Environmental Protection Strategy
  Research Center, Jiangsu University, Zhenjiang, 212013 Jiangsu, China}

\author{Lixin Tian}\email{tianlx@ujs.edu.cn}
\affiliation{School of Mathematical Science, Nanjing Normal University,
  Nanjing 210042, Jiangsu, China}
\affiliation{Energy Development and Environmental Protection Strategy
  Research Center, Jiangsu University, Zhenjiang, 212013 Jiangsu, China}

\author{H. Eugene Stanley}
\affiliation{Center for Polymer Studies and Department of Physics,
  Boston University, Boston, MA 02215, USA}

\begin{abstract}

The limited penetrable horizontal visibility graph algorithm was
recently introduced to map time series in complex networks. We extend
this visibility graph and create a directed limited penetrable
horizontal visibility graph and an image limited penetrable horizontal
visibility graph. We define the two algorithms and provide theoretical
results on the topological properties of these graphs associated with
different types of real-value series (or matrices). We perform several
numerical simulations to further check the accuracy of our theoretical
results. Finally we present an application of the directed limited
penetrable horizontal visibility graph for measuring real-value time
series irreversibility, and an application of the image limited
penetrable horizontal visibility graph that discriminates noise from
chaos. The empirical results show the effectiveness of our proposed
algorithms.

\end{abstract}

\maketitle

\section{introduction}

The complex network analysis of univariate (or multivariate) time series
has recently attracted the attention of reseachers working in a wide
range of fields \cite{1}. Over the past decade several methodologies
have been proposed for mapping a univariate and multivariate time series in
a complex network \cite{2,3,4,5,6,7,8,9}. These include constructing a
complex network from a pseudoperiodic time series \cite{2}, using a
visibility graph (VG) algorithm \cite{3}, a recurrence network (RN)
method \cite{4}, a stochastic processes method \cite{5}, a coarse
geometry theory \cite{6}, a nonlinear mutual information method
\cite{7}, event synchronization \cite{8}, and a phase-space
coarse-graining method \cite{9}. These methods have been widely used to
solve problems in a variety of research fields
\cite{10,11,12,13,14,15,16,17,18,19,20}.

Among all these time series complex network analysis algorithms,
visibility algorithms \cite{3,21,22} are the most efficient when
constructing a complex network from a time series. Visibility algorithms
are a family of rules for mapping a real-value time series on graphs
that display several cases. In all cases each time series datum is
assigned to a node, but the connection criterion differs. For example,
in the natural visibility graph (NVG) two nodes $i$ and $j$ are
connected if the geometrical criterion
$x(t_{k})<x(t_{i})+[x(t_{j})-x(t_{i})]\frac{t_{k}-t_{i}}{t_{j}-t_{k}},
\forall t_{k}\in (t_{i},t_{j})$ is fulfilled within the time series \cite{3}.  In
the parametric natural visibility graph (PNVG) case there are
three steps when using this algorithm to map a time series to a complex
network, (i) build an NVG [3] as described above using common NVG
criteria in the mapping, (ii) set the direction and angle,
$\alpha_{ij}=arctg\frac{x(t_{j})-x(t_{i})}{t_{j}-t_{i}},i<j$ for every
link of the NVG, and (iii) use the parameter view angle rule $\alpha$,
$(i,j)\in PNVG(\alpha), \alpha_{ij} < \alpha$ to select links from the
directed and weighted graph \cite{21}. In the horizontal visibility graph (HVG)
case, this algorithm is similar to the NVG algorithm but has a
modified mapping criterion. Here two nodes $i$ and $j$ are connected if
$x(t_{k})<inf(x(t_{i}),x(t_{j})), \forall t_{k}\in (t_{i},t_{j})$ \cite{22}. These
visibility algorithms have been successfully implemented in a variety of
fields \cite{23,24,25}.

Recently a limited penetrable visibility graph (LPVG) \cite{26,27} and a
multiscale limited penetrable horizontal visibility graph (MLPHVG)
\cite{28} were developed from the visibility graph (VG) and the
horizontal visibility graph (HVG) to analyze nonlinear time
series. The LPVG and MLPHVG have been successfully used to analyze a
variety of real signals across different fields, e.g., experimental flow
signals [26-27], EEG signals \cite{28,29}, and electromechanical signals
\cite{30}. Research has shown that the LPVG and MLPHVG inherit the
merits of the VG, but also successfully screen out noise, which makes
them particularly useful when analyzing signals polluted by unavoidable
noise \cite{26,27,28,29,30}.

Abundant empirical results have already been obtained using the VG
algorithm and its extensions, e.g., the PNVG \cite{21}, the HVG
\cite{22}, the LPVG \cite{26}, and the MLPHVG \cite{28}. Thus far there
has been little research focusing on rigorous theoretical results.
Recently Lacasa et al. presented topological properties of the
horizontal visibility graph associated with random time series
\cite{22}, periodic series \cite{31}, and other stochastic and chaotic
processes \cite{32}. They extended the family of visibility algorithms
to map scalar fields of an arbitrary dimension onto graphs and provided
analytical results on the topological properties of the graphs
associated with different types of real-value matrices \cite{33}. Wang
et al. \cite{34} focused on a class of general horizontal visibility
algorithms, the limited penetrable horizontal visibility graph (LPHVG),
and presented exact results on the topological properties of the limited
penetrable horizontal visibility graph associated with a random
series. Here we use the previous works \cite{22,31,32,33,34}, focus our attention on
the limited penetrable horizontal visibility graph, and present some
analytical properties.

This paper is organized as follows. In Section II of this paper we introduce the limited
penetrable horizontal visibility graph family. In Section III we derive the
analytical properties of the different versions of associated limited
penetrable horizontal visibility graphs of a generic random time series
(or a random matrix) and present several numerical simulations to check
their accuracy. In Section IV we show some applications of the directed
limited penetrable horizontal visibility graph and the image limited
penetrable horizontal visibility graph. In Section V we present our conclusions.

\section{limited penetrable horizontal visibility graph family}

The LPHVG algorithm \cite{28,34} and its extensions are called the LPHVG
family. We here present three versions of the LPHVG algorithm, the
limited penetrable horizontal visibility graph, LPHVG$(\rho)$, the
directed limited penetrable horizontal visibility graph, DLPHVG$(\rho)$,
and the image limited penetrable horizontal visibility graph of order
$n$, ILPHVG$_{n}(\rho)$.

\subsection{Limited Penetrable Horizontal Visibility Graph [LPHVG$(\rho)$]}

The limited penetrable horizontal visibility graph [LPHVG$(\rho)$]
\cite{34} is a geometrically simpler and analytically solvable version
of VG \cite{3}, LPVG \cite{30}, and MLPHVG \cite{28}. To
define it we let $\{x_{i}\}_{i=1,2,...,N}$ be a time series of $N$ real
numbers. We set the limited penetrable distance to $\rho$, and
LPHVG$(\rho)$ maps the time series on a graph with $N$ nodes and an
adjacency matrix $\textbf{A}$. Nodes $x_{i}$ and $x_{j}$ are connected
through an undirected edge ($A_{ij} = A_{ji} = 1$) if $x_{i}$ and
$x_{j}$ have a limited penetrable horizontal visibility (see Fig.~1),
i.e., if $\rho \geq 0$ intermediate data $x_{q}$ follows
\begin{equation}\label{eq7}
x_{q}\geq inf\{x_{i},x_{j}\},\forall q\in (i,j),\aleph (q) \leq \rho,
\end{equation}
where $\aleph(q)$ is the number of $q$. The graph spanned by this
mapping is the limited penetrable horizontal visibility graph
[LPHVG$(\rho)$]. When we set the limited penetrable distance $\rho$ to
0, then LPHVG(0) degenerates into an HVG \cite{22}, i.e.,
LPHVG(0)~=~HVG. When $\rho\neq 0$ there are more connections between any
two LPHVG$(\rho)$ nodes than in HVG.  Fig.~1(b) shows the new
established connections (red lines) when we infer the LPHVG(1) using
HVG. Note that the LPHVG$(\rho)$ of a time series has all the properties of its
corresponding HVG, e.g., it is connected and invariant under affine
transformations of series data \cite{22}.

\begin{figure}[H]
\centering \scalebox{0.6}[0.6]{\includegraphics{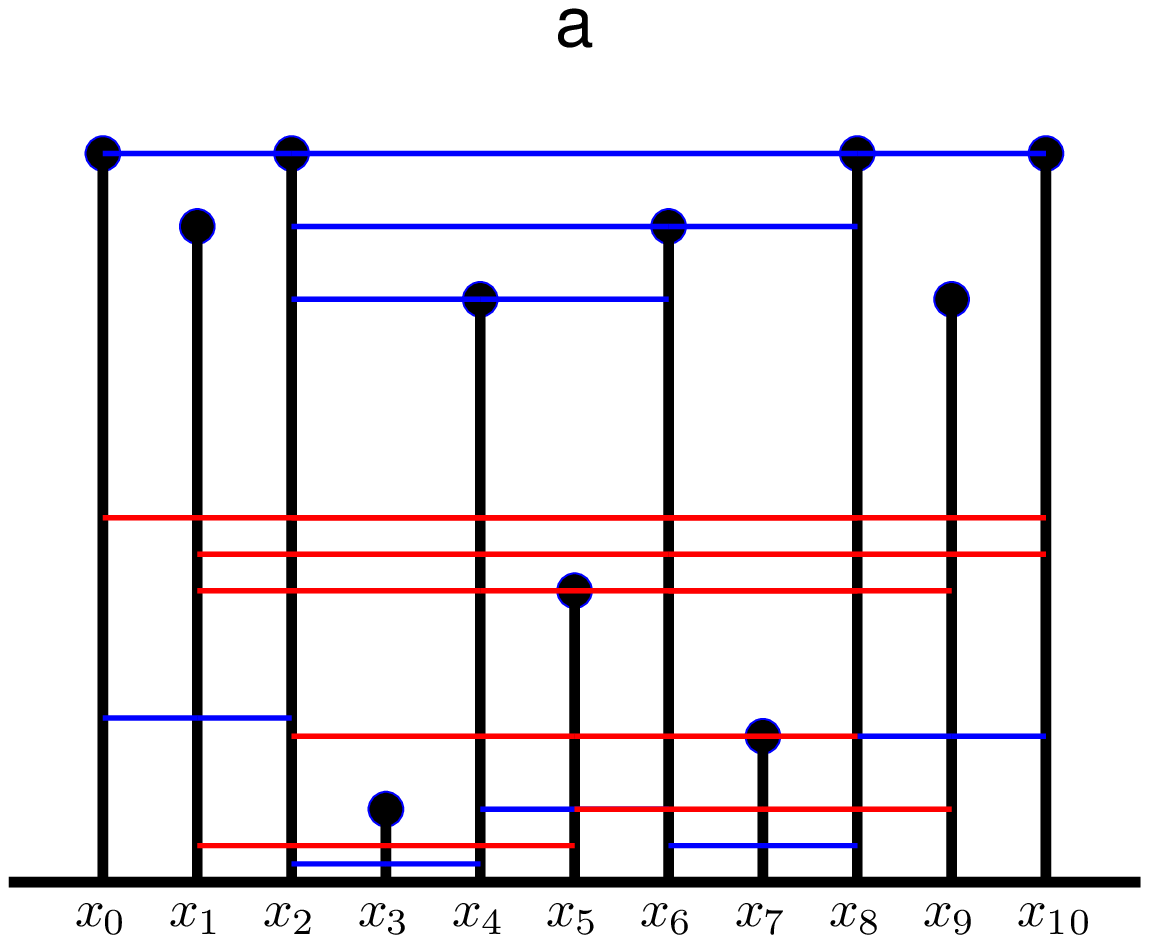}}
\scalebox{0.6}[0.6]{\includegraphics{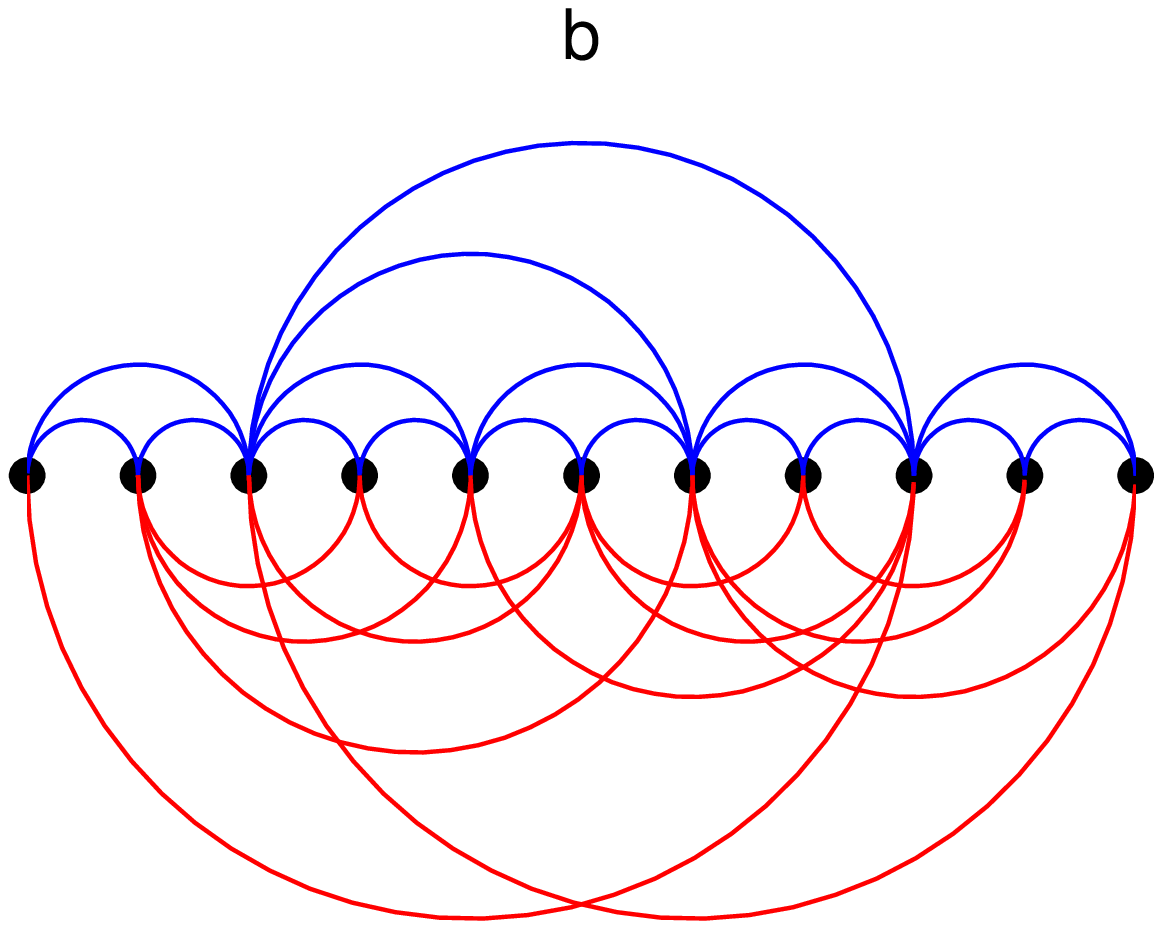}}
\end{figure}
\begin{figure}[H]
\caption{Example of (a) a time series (11 data values) and (b) its corresponding LPHVG(1), where every node corresponds to time
    series data in the same order. The horizontal penetrable visibility
    lines between data points define the links connecting nodes in the
    graph.}
\end{figure}

\subsection{Directed limited penetrable horizontal visibility graph
  [DLPHVG$(\rho)$]}

The limited penetrable horizontal visibility graph
[LPHVG$(\rho)$] is undirected, because penetrable visibility does not
have a predefined temporal arrow. Directionality can be added by using
directed networks. Here we address the directed version and define a
directed limited penetrable horizontal visibility graph
[DLPHVG$(\rho)$], where the degree $k(x_{t})$ of the node $x_{t}$ is
split between an ingoing degree $k_{\rm in}(x_{t})$ and an outgoing
degree $k_{\rm out}(x_{t})$ such that $k(x_t)=k_{\rm in}(x_t)+k_{\rm
  out}(x_t)$. We define the ingoing degree $k_{\rm in}(x_t)$ to be the
number of links of node $x_t$ with past nodes associated with data in
the series, i.e., nodes with $t' < t$. Conversely, we define the
outgoing degree $k_{\rm out}(x_t)$ to be the number of links with future
nodes, i.e., nodes with $t'' > t$. Thus DLPHVG$(\rho)$ maps the time
series into a graph with $N$ nodes and an adjacency matrix
$\textbf{A}=\textbf{A}_{\rm in}+\textbf{A}_{\rm out}$, where
$\textbf{A}_{\rm in}$ is a lower triangular matrix and $\textbf{A}_{\rm
  out}$ is a upper triangular matrix. Nodes $x_{t'}$ and $x_t$, $t'<t$
(or $x_t$ and $x_{t''}$, $t<t''$) are connected through a directed edge
$x_{t'} \rightarrow x_t$, i.e., $A_{t't}=1$ (or $x_t \rightarrow
x_{t''}$, i.e. $A_{tt''}=1$) if it satisfies Eq.~(1).

Fig.~2 shows a graphical representation of the definition. As in the
degree distribution $P(k)$, we use the ingoing and outgoing degree
distributions of a DLPHVG$(\rho)$ to define the probability
distributions of $k_{\rm out}$ and $k_{\rm in}$ on the graph, which are
$P_{\rm out}(k) \equiv P(k_{\rm out} = k)$ and $P_{\rm in}(k) \equiv
P(k_{\rm in} = k)$, respectively. We see the asymmetry of the resulting
graph in a first approximation when we use the invariance of the
outgoing (or ingoing) degree series under a time reversal.

\begin{figure}[H]
\centering \scalebox{0.6}[0.6]{\includegraphics{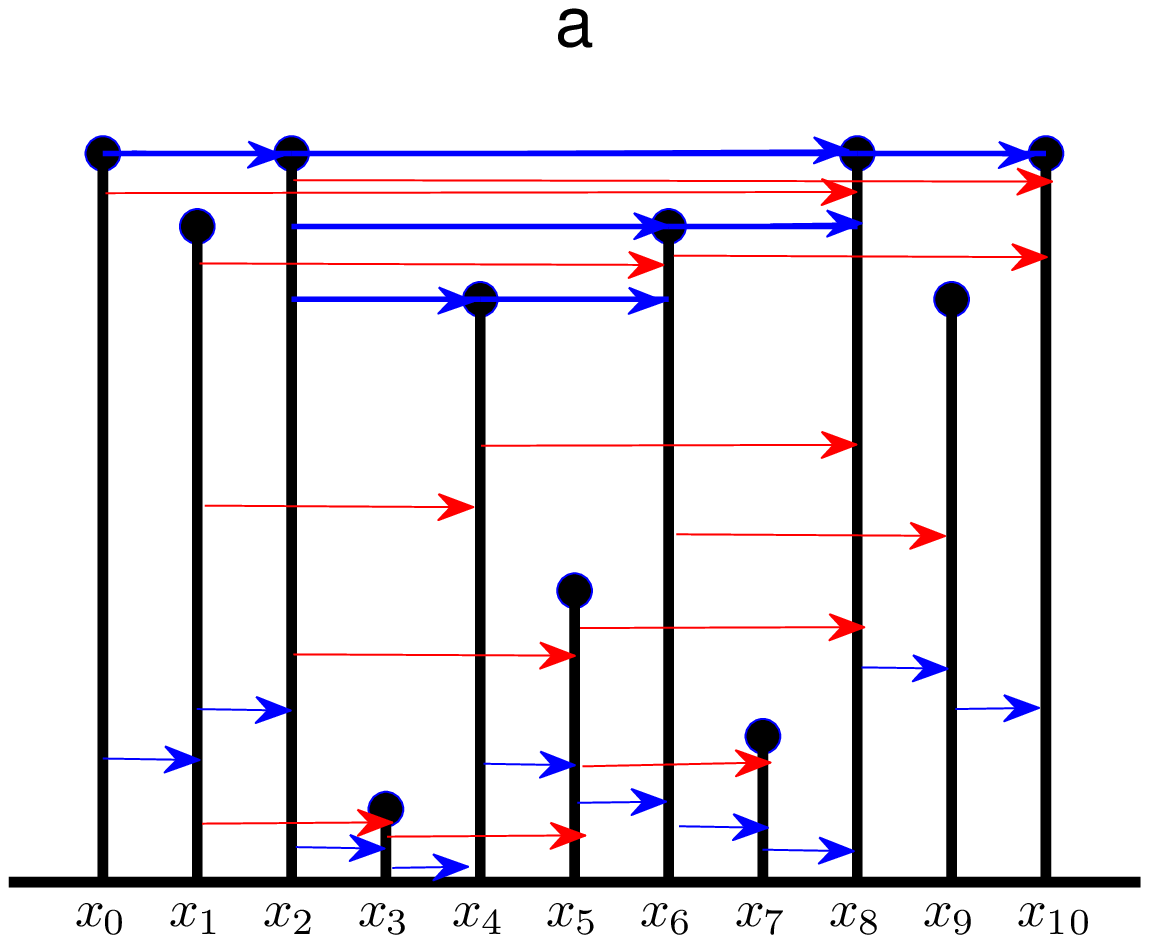}}
\scalebox{0.6}[0.6]{\includegraphics{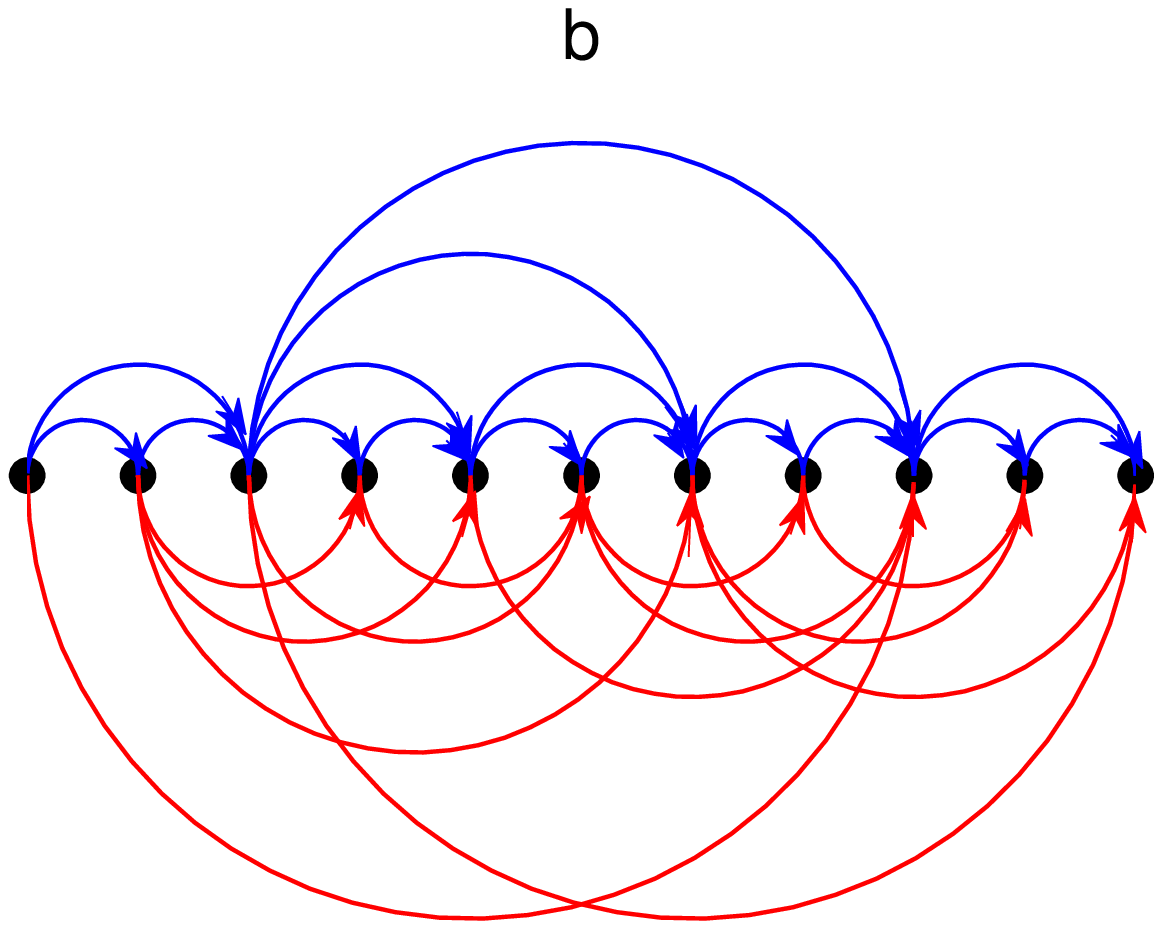}}
\end{figure}
\begin{figure}[H]
\caption{Graphical illustration of DLPHVG(1). (a) Plot of a
    sample time series $\{x_t\},t=0,1,2,...,10$. Each datum in the
    series is mapped to a node in the graph. Arrows, describing allowed
    directed penetrable visibility, link nodes. (b) Plot of the associated
    DLPHVG(1). In this graph, each node has an ingoing degree $k_{\rm
      in}$, which accounts for the number of links with past nodes, and
    an outgoing degree $k_{\rm out}$, which in turn accounts for the
    number of links with future nodes.}
\end{figure}

\subsection{Image limited penetrable horizontal visibility graph of order
  $n$ [ILPHVG$_{n}(\rho)$]}

One-dimensional versions of the limited
penetrable horizontal visibility graph [LPHVG$(\rho)$] and directed
limited penetrable horizontal visibility graph [DLPHVG$(\rho)$] are used
to map landscapes (time series) on complex networks. As in the
definition of IVG$_{n}$ \cite{33}, the definition of LPHVG$(\rho)$ can
also be extended and applied to two-dimensional manifolds by extending
the LPHVG$(\rho)$ criteria of Eq.~(1) along one-dimensional sections of
the manifold. To define the image limited penetrable horizontal
visibility graph of order $n$ [ILPHVG$_{n}(\rho)$] we let $\textbf{X}$
be a $N \times N$ matrix for an arbitrary entry $(i,j)$ and partition the
plane into $n$ directions such that direction $p$ is at an angle with
the row axis of $2\pi(p-1)/n$, where $p=1,2,...,n$. The image limited
penetrable visibility graph of order $n$, ILPHVG$_{n}(\rho)$, has
$N^{2}$ nodes, each of which is labeled using a duple $(i,j)$ associated
with the entry indices $x_{ij}$, such that two nodes, $x_{ij}$ and
$x_{i'j'}$, are linked when (i) $x_{i'j'}$ belongs to one of the $n$
angular partition lines, and (ii) $x_{ij}$ and $x_{i'j'}$ are linked in
the LPHVG$(\rho)$ defined over the ordered sequence that includes $(i,j)$
and $(i',j')$. For example, in ILPHVG$_{4}(1)$ the penetrable visibility
between two points $x_{ij}$ and $x_{i'j'}$ is
\begin{equation}\label{eq1}
\begin{array}{l}
i = i',x_{iq} \geq inf\{x_{ij},x_{i'j'}\},\forall q\in
(j,j'),\aleph(q)\leq \rho,
\end{array}
\end{equation}
or
\begin{equation}\label{eq1}
\begin{array}{l}
j = j',x_{qj} \geq inf\{x_{ij},x_{i'j'}\}, \forall q\in
(i,i'),\aleph(q)\leq \rho.
\end{array}
\end{equation}

Fig.~3(a) shows a sample matrix in which $x_{0}$ is the central entry,
which shows the ILPHVG$_{4}$(1) algorithm evaluated along the vertical
and horizontal directions. Fig.~3(b) shows the connectivity pattern associated to the entry $x_{0}$
of the ILPHVG$_{4}$(1) algorithm. Fig.~3(c) shows the
ILPHVG$_{8}$(1) algorithm evaluated along the vertical, horizontal, and
diagonal directions. Fig.~3(d) shows the connectivity pattern associated to the entry $x_{0}$ of the
ILPHVG$_{8}$(1) algorithm.

\begin{figure}[H]
\centering \scalebox{0.6}[0.6]{\includegraphics{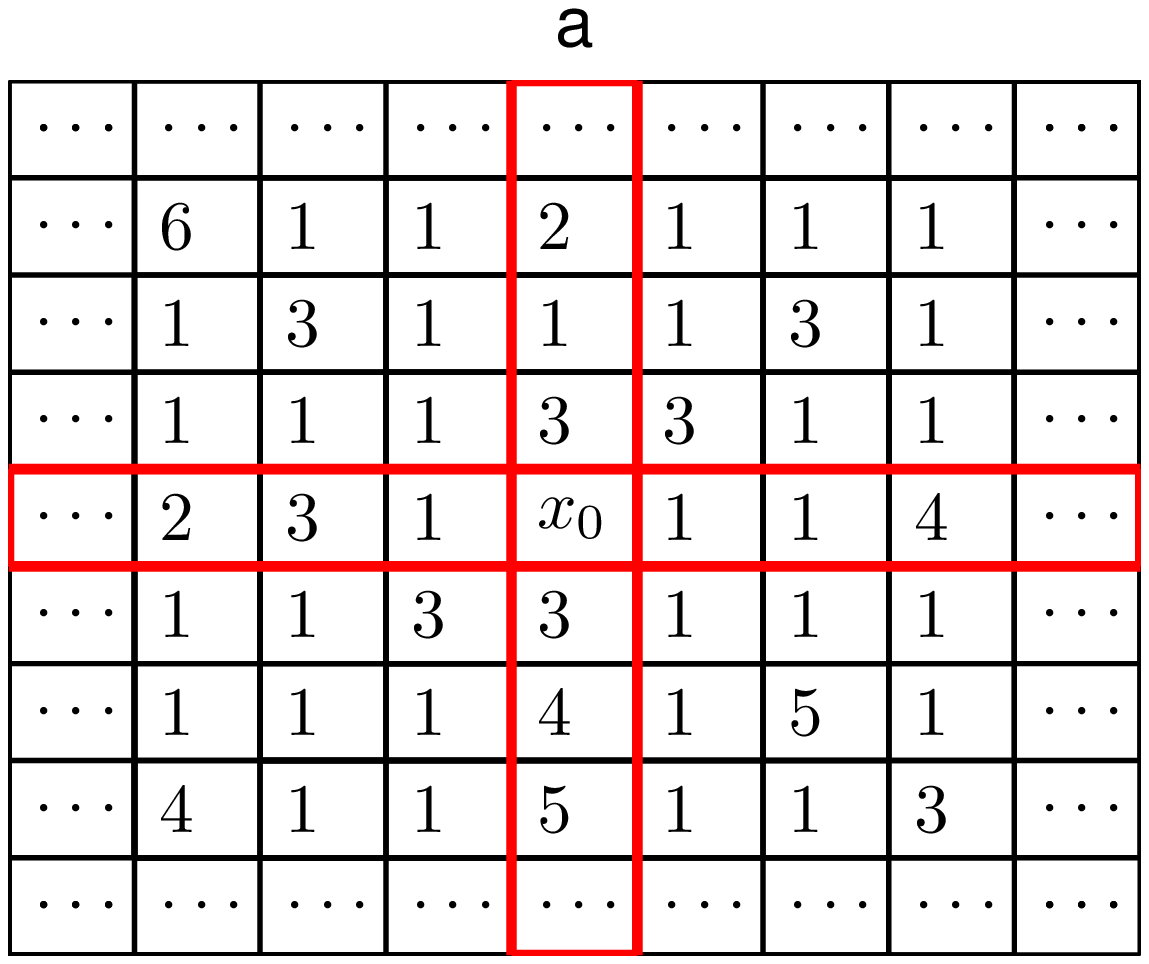}}
\scalebox{0.6}[0.6]{\includegraphics{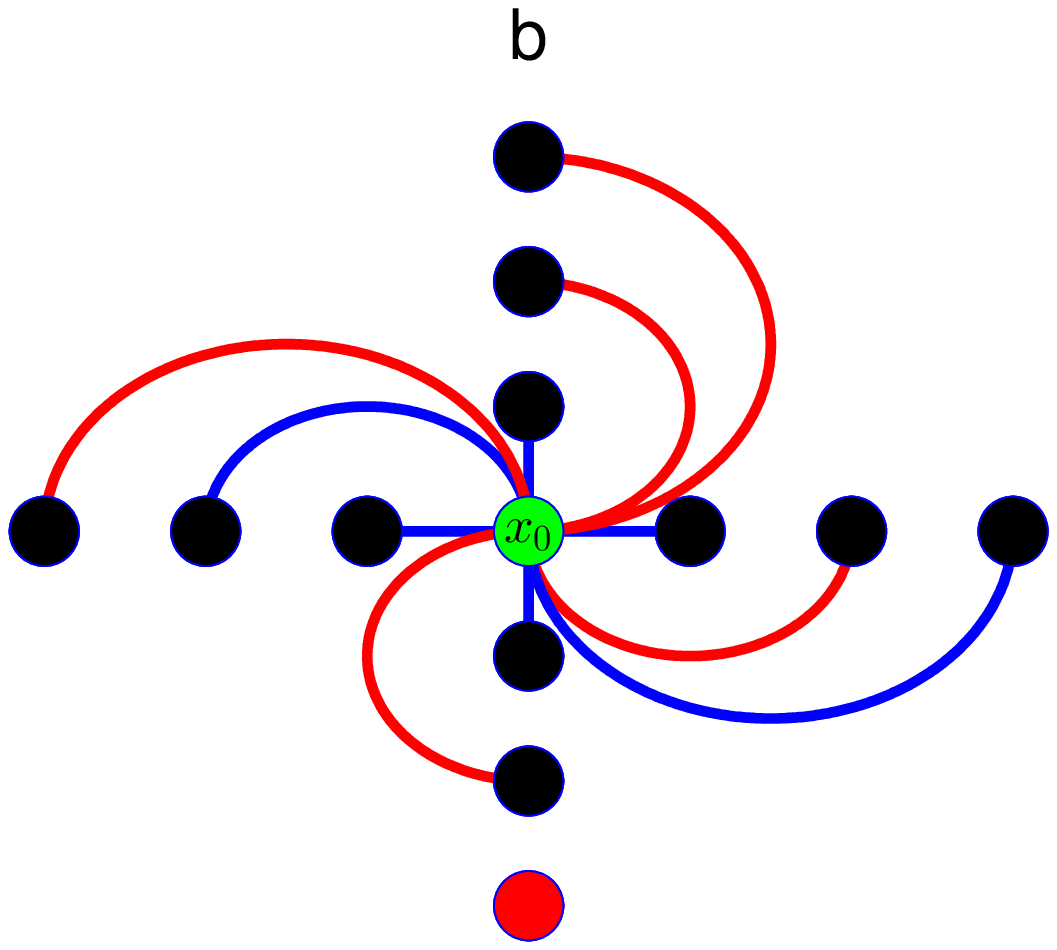}}
\scalebox{0.6}[0.6]{\includegraphics{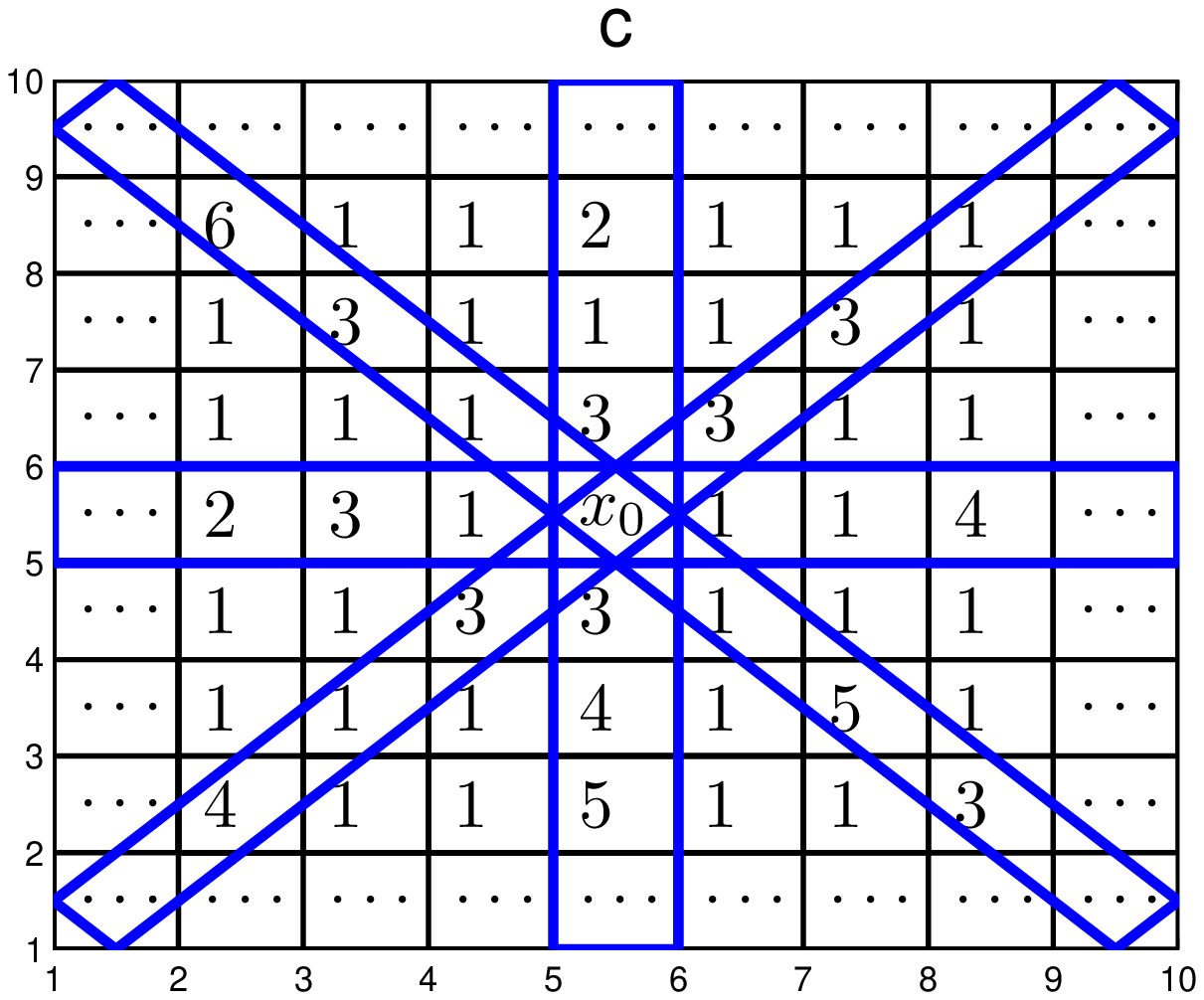}}
\scalebox{0.6}[0.6]{\includegraphics{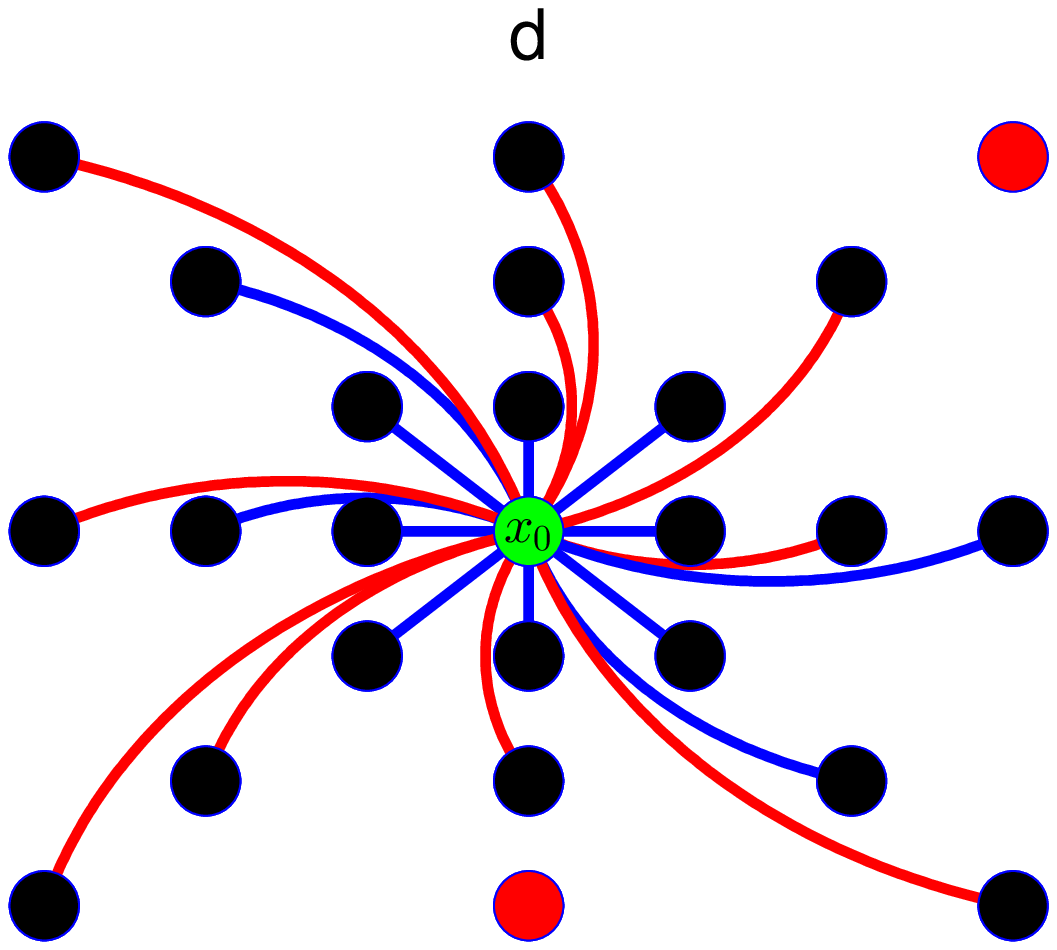}}
\end{figure}
\begin{figure}[H]
\caption{Graphical illustration of ILPHVG$_{n}(\rho)$. In
    Fig. 3(a) we depict a sample matrix where $x_{0}=2$ is the central
    entry, which shows the ILPHVG$_{4}$(1) algorithm is evaluated along
    the vertical and horizontal directions, and in Fig. 3(b) of the same
    figure we describe the connectivity pattern associated to this entry
    $x_{0}$ in the case of ILPHVG$_{4}$(1). Fig. 3(c) shows the
    ILPHVG$_{8}$(1) algorithm is evaluated both along the vertical and
    horizontal directions and along the diagonals directions, and in
    Fig. 3(d) of the same figure we describe the connectivity pattern
    associated to this entry $x_{0}$.}
\end{figure}

\section{Theoretical results on the topological properties}

\noindent
\textbf{Theorem 1. \cite{34}} If we let $X(t)$ be a bi-infinite sequence
of $i.i.d.$, a random variable $X$ with probability density $f(x)$, then
the degree distribution of its associated LPHVG$(\rho)$ is
\begin{equation*}\label{eq1}
\begin{array}{l}
P(k)=
\begin{cases}
\frac{1}{2\rho+3}(\frac{2\rho+2}{2\rho+3})^{k-2(\rho+1)},k\geq 2\rho+2.\\
0, otherwise.
\end{cases}
\end{array}
\end{equation*}
The mean degree $\langle k\rangle$ is
\begin{equation*}\label{eq1}
\begin{array}{l}
\langle k\rangle = 4(\rho+1).
\end{array}
\end{equation*}
Reference \cite{34} [Wang et al., 2017] provides a lengthy proof of this
theorem. We here propose an alternative shorter proof.

\textbf{Proof.}  We let $x$ be an arbitrary datum of the $i.i.d.$ random
time series. The probability that its limited penetrable horizontal
visibility is interrupted by two bounding data, one datum $x_{bl}$ on
its left and one $x_{br}$ on its right. There are $2\rho$ penetrable
data that are larger than $x$ between the two bounding data, $\rho$
penetrable data $x_{pl}^{1},x_{pl}^{2},...,x_{pl}^{\rho}$ on the left
and $\rho$ data $x_{pr}^{1},x_{pr}^{2},...,x_{pr}^{\rho}$ on the right
of $x$. These $2\rho+2$ data are independent of $f(x)$, then
\begin{equation}\label{eq1}
\begin{array}{l}
\Phi_{2\rho+2} = \int_{-\infty}^{\infty}\int_{x}^{\infty}
\int_{x}^{\infty}\int_{x}^{\infty}...\int_{x}^{\infty}
\int_{x}^{\infty}...\int_{x}^{\infty}f(x)f(x_{bl})
f(x_{br})f(x_{pl}^{1})...f(x_{pl}^{\rho})
f(x_{pr}^{1})...f(x_{pr}^{\rho})dx_{pr}^{\rho}...\\
...dx_{pr}^{1}dx_{pl}^{\rho}...dx_{pl}^{1}dx_{br}dx_{bl}dx.
\end{array}
\end{equation}

We define the cumulative probability distribution function $F(x)$ of any
probability distribution $f(x)$ to be
\begin{equation}\label{eq1}
\begin{array}{l}
F(x) = \int_{-\infty}^{x}f(t)dt.
\end{array}
\end{equation}
Then we rewrite Eq.~(4) to be
\begin{equation}\label{eq1}
\begin{array}{l}
\Phi_{2\rho+2} =\int_{-\infty}^{\infty}f(x)[1-F(x)]^{2\rho+2}dx=\frac{1}{2\rho+3}.
\end{array}
\end{equation}
The probability $P(k)$ that the datum penetrates no more than $\rho$
time seeing $k$ data is
\begin{equation}\label{eq1}
\begin{array}{l}
P(k) = \Phi(k)\Phi_{2\rho+2}=\frac{1}{2\rho+3}\Phi(k),
\end{array}
\end{equation}
where $\Phi(k)$ is the probability that datum $x$ penetrates no more
than $\rho$ time seeing at least $k$ data. We can recurrently calculate
$\Phi(k)$ to be
\begin{equation}\label{eq1}
\begin{array}{l}
\Phi(k)=\Phi(k-1)(1-\Phi_{2\rho+2})=\frac{2\rho+2}{2\rho+3}\Phi(k-1),
\Phi(2\rho+2)=1,
\end{array}
\end{equation}
from which we deduce
\begin{equation}\label{eq1}
\begin{array}{l}
\Phi(k)=(\frac{2\rho+2}{2\rho+3})^{k-2(\rho+1)}\Phi(2\rho+2)
=(\frac{2\rho+2}{2\rho+3})^{k-2(\rho+1)}.
\end{array}
\end{equation}
Thus we finally obtain
\begin{equation}\label{eq1}
\begin{array}{l}
P(k) =
\begin{cases}
\Phi(k)\Phi_{2\rho+2}=\frac{1}{2\rho+3}(\frac{2\rho+2}{2\rho+3})^{k-2(\rho+1)},k
\geq 2\rho+2,\\
0,otherwise,
\end{cases}
\rho=0,1,2,...
\end{array}
\end{equation}
Then the mean degree $\langle k\rangle$ of the limited penetrable
horizontal visibility graph associated to an uncorrelated random process
is
\begin{equation}\label{eq1}
\begin{array}{l}
\langle k\rangle = \sum\limits_{k=2\rho+2}^{\infty} kP(k) =
\sum\limits_{k=2\rho+2}^{\infty}\frac{k}{2\rho+3}
(\frac{2\rho+2}{2\rho+3})^{k-2(\rho+1)}=4(\rho+1).
\end{array}
\end{equation}
Theorem 1 shows the exact degree distribution for LPHVG$(\rho)$, which
indicates that the degree distribution $P(k)$ of LPHVG$(\rho)$
associated to $i.i.d.$ random time series has a unified exponential
form, independent of the probability distribution from which the series
was generated.

\textbf{Theorem 2.} We let $X(t)$ be a bi-infinite sequence of $i.i.d.$, a
random variable $X$ with probability density $f(x)$, and consider a limited
penetrable horizontal visibility graph associated with $X(t)$. We let
$\langle K(x)\rangle$ be a mean degree of the node associated with a datum of height
$x$ and define it
$$\langle K(x)\rangle = 2(\rho+1)-2(\rho+1)ln(1-F(x)),F(x) =
\int_{-\infty}^{x}f(t)dt. $$

\textbf{Proof.} We define $P(k|x)$ to be the conditional probability
that a given node has degree $k$ when its height is $x$. Using the
constructive proof process of $P(k)$ in Ref.~\cite{34} [Wang et al.,
  2017], we calculate $P(k|x)$ to be
\begin{equation}\label{eq1}
\begin{array}{l}
P(k|x) =
\sum\limits_{h=0}^{k-2(\rho+1)}(2\rho+1)^{h}
\frac{(-1)^{k-2(\rho+1)}}{h![k-2(\rho+1)-h]!}[1-F(x)]^{2(\rho+1)}
[ln(1-F(x))]^{k-2(\rho+1)}\\
=[1-F(x)]^{2(\rho+1)}[2(\rho+1)ln(1-F(x))]^{k-2(\rho+1)}
\frac{(-1)^{k-2(\rho+1)}}{[k-2(\rho+1)]!}.
\end{array}
\end{equation}
Then $\langle K(x)\rangle$ is
\begin{equation}\label{eq1}
\begin{array}{l}
\langle K(x)\rangle = \sum\limits_{k=2(\rho+1)}^{\infty}kP(k|x).
\end{array}
\end{equation}
We let $k-2(\rho+1) = \alpha$, $2(\rho+1)ln[1-F(x)]=A$ and deduce
\begin{equation}\label{eq1}
\begin{array}{l}
\langle K(x)\rangle =
2(\rho+1)[1-F(x)]^{2(\rho+1)}\sum\limits_{\alpha=0}^{\infty}
\frac{(-1)^{\alpha}A^{\alpha}}{\alpha!}+[1-F(x)]^{2(\rho+1)}
\sum\limits_{\alpha=1}^{\infty}\frac{(-1)^{\alpha}A^{\alpha}}{(\alpha-1)!}\\
=2(\rho+1)-A = 2(\rho+1)-2(\rho+1)ln[1-F(x)].
\end{array}
\end{equation}

Theorem 2 shows the relation between data height $x$ and the mean degree
of the nodes associated with the data of height $x$. The result
indicates that the $\langle K(x)\rangle$ is a monotonically increasing
function of $x$. Thus we conclude that the hubs of LPHVG$(\rho)$ are the
data with largest values. We check the accuracy of the result within
finite series. Fig.~ 4(a) shows a plot of the numerical values of
$\langle K(x)\rangle$ of LPHVG($\rho$), $\rho=0,2,4,6,8,10$ associated
with the random series of 1000 data extracted from a uniform
distribution when $F(x)=x$. The theoretical results (red lines) show a
perfect agreement [Eq.~(14)]. To check the finite size effect, Fig.~4(b)
shows a plot of the numerical values of $\langle K(x)\rangle$ of
LPHVG(2) associated with random series of 500, 1000, 1500,
2000 data. We use root mean square error (RMSE) to measure the agreement
between the numerical and theoretical results. We find that when the
size $N$ of the time series increases, the RMSE between the numerical
and theoretical results decreases, indicating an increase in agreement.

\begin{figure}[H]
\centering \scalebox{0.6}[0.6]{\includegraphics{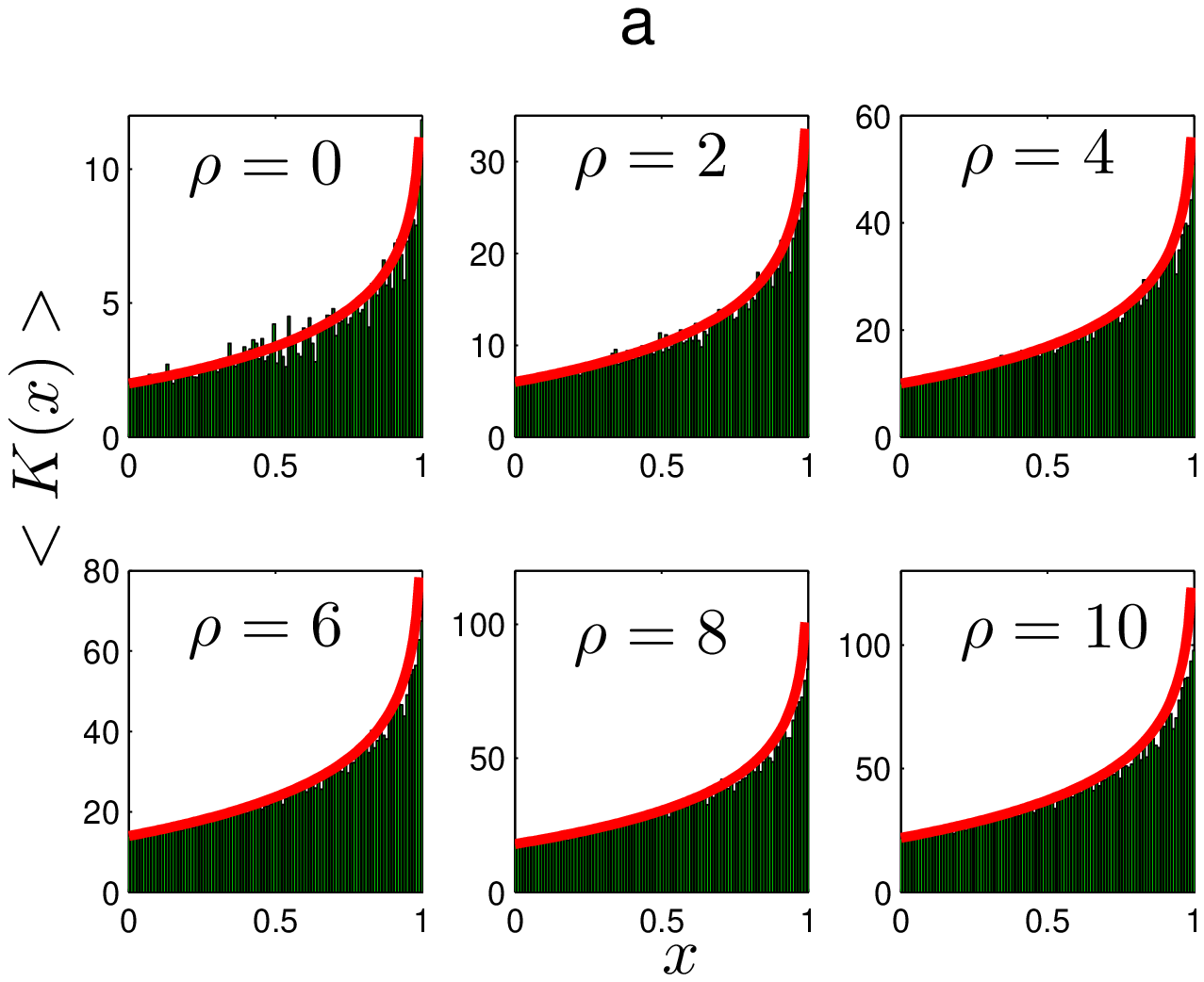}}
\scalebox{0.6}[0.6]{\includegraphics{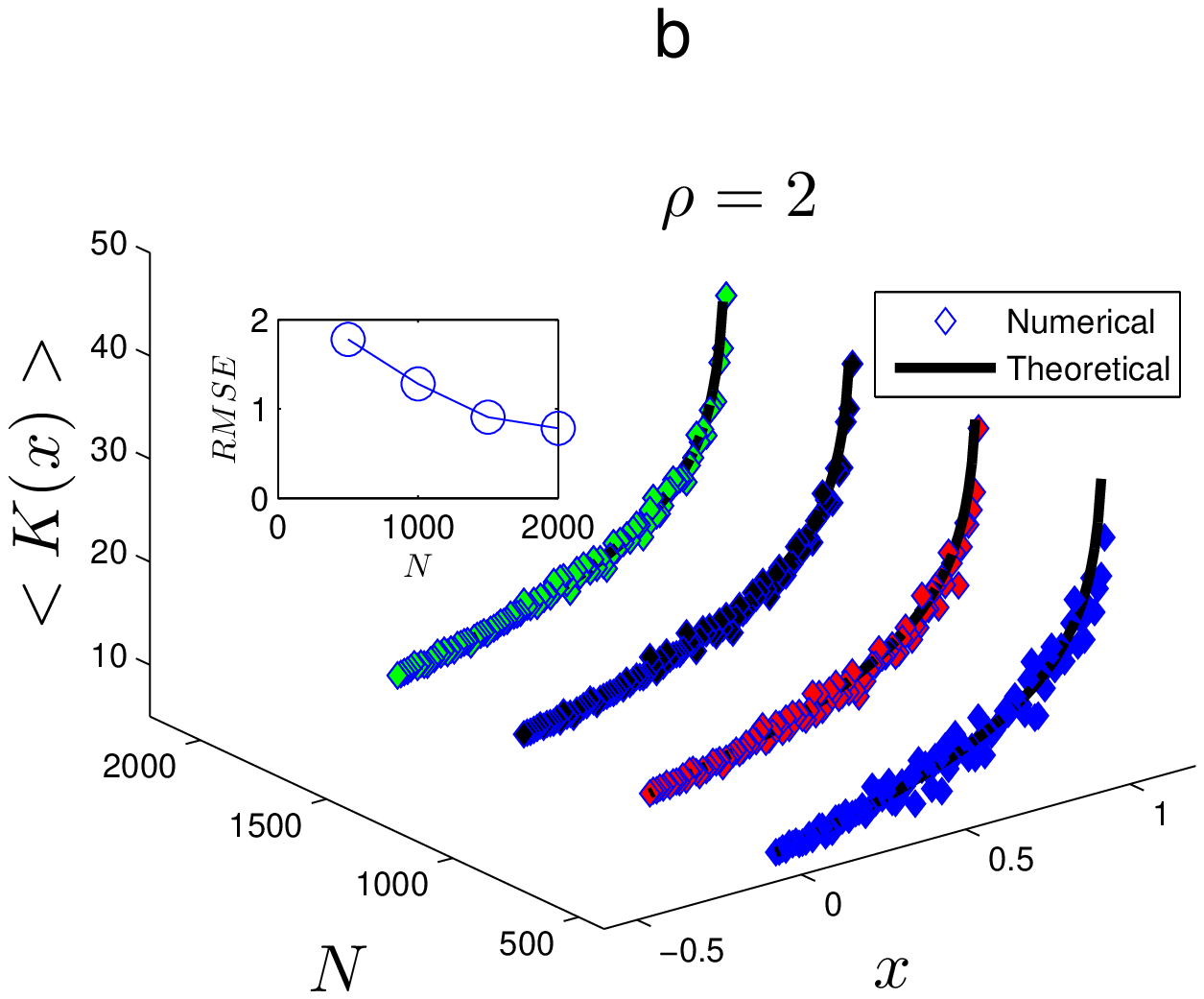}}
\end{figure}
\begin{figure}[H]
\caption{(a) The relation between data height $x$ and the node
    degree $\langle K(x)\rangle$ under different penetrable distance
    $\rho$. (b) the relation between data height $x$ and the node degree
    $\langle K(x)\rangle$ under different time series size $N$.}
\end{figure}

\textbf{Theorem 3.}  We let $X_{t}$ be an infinite periodic series of
period $T$ with no repeated values within a period. The normalized mean
distance $\langle d\rangle$ of LPHVG$(\rho)$ associated with $X_{t}$ is
$$\langle d\rangle \sim [4(\rho+1)-\langle k(T)\rangle],$$
where
$$\langle k(T)\rangle=4(\rho+1)(1-\frac{2\rho+1}{2T}),\rho \ll T.$$

\textbf{Proof.} To calculate $\langle k(T)\rangle$ we consider an
infinite periodic series of period $T$ with no repeated values in a
period and denote it
$X_{t}=\{...,x_{0},x_{1},x_{2},...,x_{T},x_{1},x_{2},...\},x_{0} =
x_{T}$. We let $\rho \ll T$ for the subseries $\tilde{X}_{t} =
\{x_{0},x_{1},x_{2},...,x_{T}\}$ and without losing generality assume
that $x_{0} = x_{T}$ corresponds to the largest value of the subseries,
$x_{1},..., x_{\rho}, x_{T-\rho},...x_{T-1}$, and corresponds to the 2nd
to $(2\rho+1)$nd largest value of the subseries. Thus we construct the
LPHVG$(\rho)$ associated with subseries $\tilde{X_{t}}$. We assume that
LPHVG$(\rho)$ has $E$ links and let $x_{i}$ be the smallest datum of the
subseries $\tilde{X}_{t}$. Because no data repetitions are allowed in
$\tilde{X}_{t}$, the degree of $x_{i}$ is $2(\rho+1)$ when constructed
from LPHVG$(\rho)$. We now remove node $x_{i}$ and its $2(\rho+1)$ links
from LPHVG$(\rho)$. The resulting graph now has $E-2(\rho+1)$ links and
$T$ nodes. We iterate this operation $T-(2\rho+1)$ times. The resulting
graph has $2(\rho+1)$ nodes, i.e., $x_{0},x_{1},...,x_{\rho},
x_{T-\rho},...x_{T-1}, x_{T}$. When these $2(\rho+1)$ nodes are
connected by $E_{r} = \binom{2\rho+2}{2}$ links, the total number of
deleted links are $E_{d} = 2(\rho+1)[T-(2\rho+1)]$. Thus the mean degree
of a limited penetrable horizontal visibility graph associated with
$X_{t}$ is
\begin{equation}\label{eq1}
\langle k(T)\rangle =
2\frac{E_{d}+E_{r}}{T}=\frac{2[(2(\rho+1))(T-(2\rho+1))+(\rho+1)(2\rho+1)]}{T}
=4(\rho+1)(1-\frac{2\rho+1}{2T})),\rho
\ll T.
\end{equation}

We let $\langle d\rangle$ be the mean distance of LPHVG$(\rho)$, $N$ be
the number of nodes, and the normalized mean distance $\langle d\rangle$
be $\langle d\rangle=\frac{\langle d\rangle}{N}$. Note that $\langle
d\rangle$ depends on $T$ for HVG associated with periodic orbits
$\langle d\rangle \sim T^{-1}$ for $N \rightarrow \infty$ [31]. Thus we
deduce that $\langle d\rangle \sim T^{-1}$ for LPHVG$(\rho)$. Using
Eq.~(15) we obtain $T^{-1} \sim [4(\rho+1)-\langle k(T)\rangle]$, and
finally obtain
\begin{equation}\label{eq1}
\langle d\rangle \sim [4(\rho+1)-\langle k(T)\rangle].
\end{equation}

This result holds for every periodic or aperiodic series ($T\rightarrow
\infty$), independent of the deterministic process that generates them,
because the only constraint in its derivation is that data within a
period not be repeated. Note that one consequence of Eq.~(15) is that
each time series has an associated LPHVG$(\rho)$ with a maximum mean
degree (for a aperiodic series) of $\langle k(\infty)\rangle = 4(\rho+1)$, which agrees
with the previous result in Eq.~(11). In Eq.~(16) the limiting solution
$\langle k(T)\rangle \rightarrow 4(\rho+1), \langle d\rangle \rightarrow
0$ holds for all aperiodic, chaotic, and random series. To check the
accuracy of the analytical result, we generate four periodic time series
($T=50$, 100, 200, and 250) with 2000 data points. The data in each
period is from the logistic map in which $\mu=4$. We construct the
limited penetrable horizontal visibility graphs with penetrable distance
$\rho = 0,1,2,...,10$ associated with this periodic time series.
Fig.~5(a) shows a plot of the mean degree of the resulting
LPHVG$(\rho)$ values with different $\rho$ values that indicate a good
agreement with the theoretical results in Eq.~(15). Fig.~5(b) shows a
calculation of the normalized mean distance $\langle d\rangle$ of
LPHVG$(\rho)$ values with $\rho=0$, 1, and 2 associated with the period
time series of $T=100,200,...,1000$. Numerical values of the mean
normalized distance $\langle d\rangle$ as a function of mean degree
$\langle k(T)\rangle$ agrees with the theoretical linear relation of
Eq.~(16).

\begin{figure}[H]
\centering \scalebox{0.6}[0.6]{\includegraphics{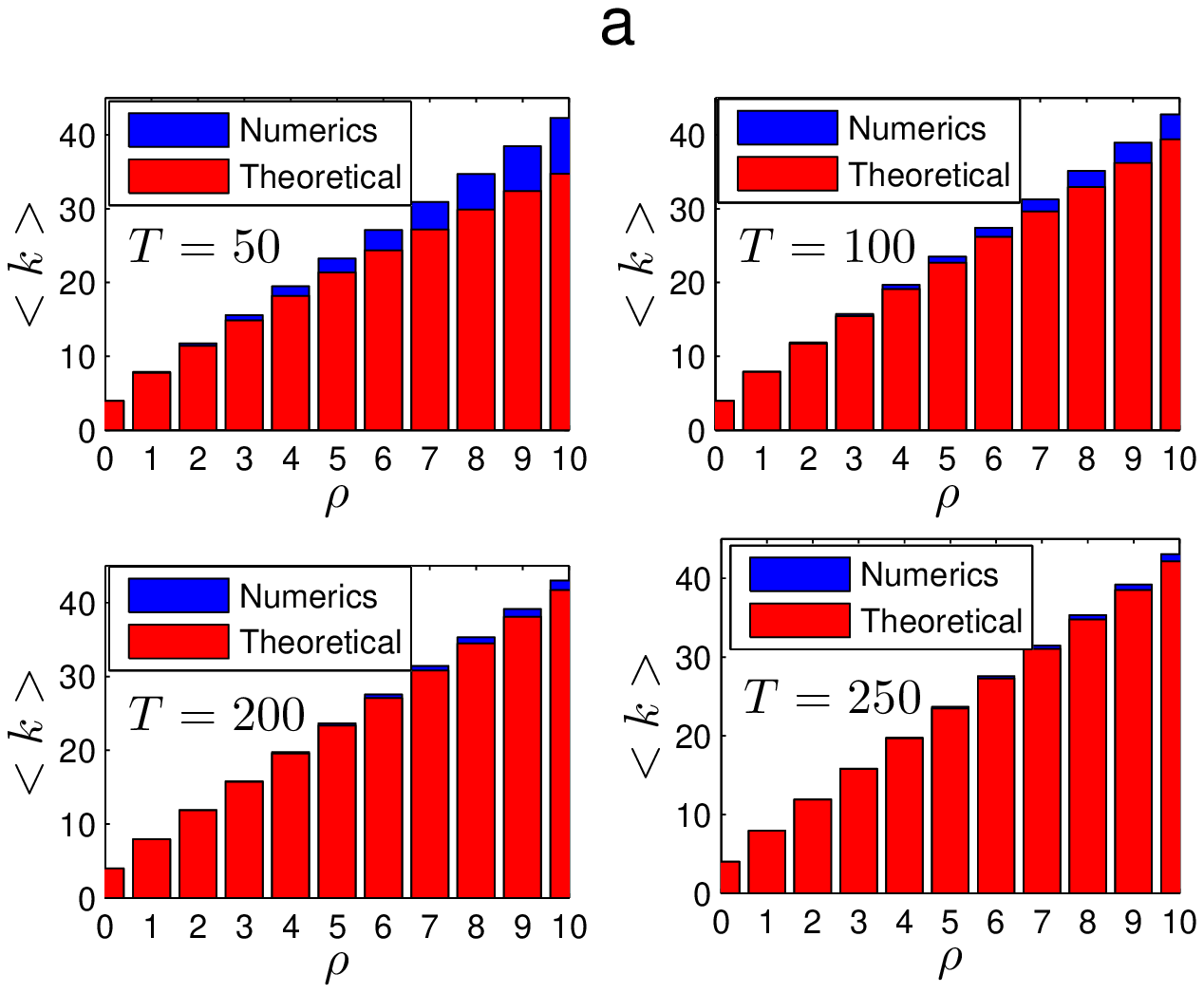}}
\scalebox{0.6}[0.6]{\includegraphics{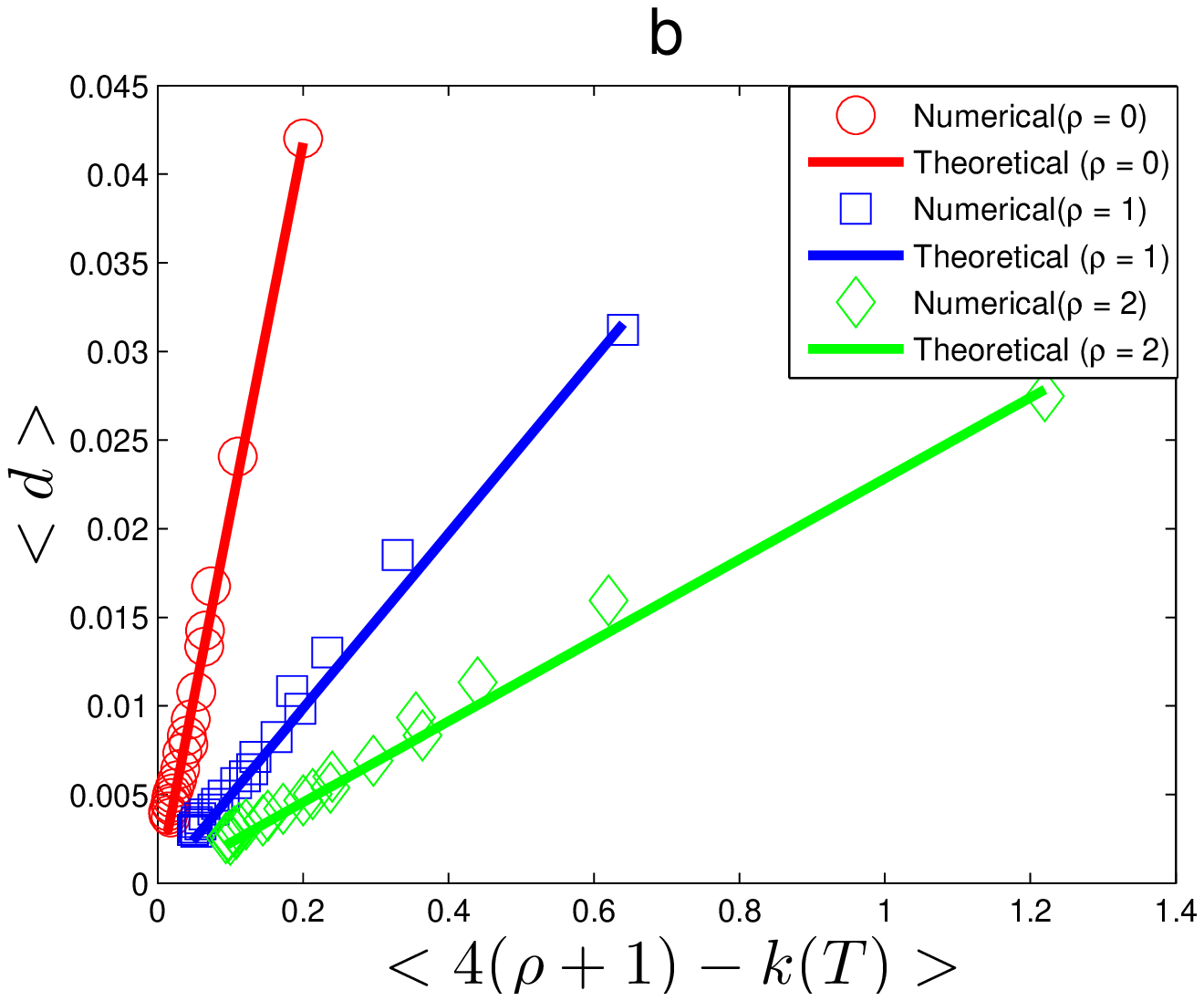}}
\end{figure}
\begin{figure}[H]
\caption{( a ) Numerical result of Eq. (15), we simulated the
    period time series ( $T=50, 100, 200, 250$ respectively ) with 2000
    data points are generated ( the data in each period from the
    Logistic map with $\mu=4$ ), ( b ) Numerical result of Eq. (16), we
    simulated the period time series with $T=100,200,...,1000$. }
\end{figure}

\textbf{Theorem 4. \cite{34}} We let $X(t)$ be a real value bi-infinite
time series of $i.i.d.$ random variables $X$ with probability
distribution $f(x)$ and examine its associated LPHVG$(\rho)$. The local
clustering coefficient distribution is then
\begin{equation*}
\begin{array}{l}
P(C_{\rm min}) =
\frac{1}{2\rho+3}exp\{[\frac{\varphi+\sqrt{\varphi^2-8C_{\rm
        min}(2\rho+1)}}{2C_{\rm
      min}}-2(\rho+1)]ln(\frac{2\rho+2}{2\rho+3})\}
\end{array}
\end{equation*}
and
\begin{equation*}
\begin{array}{l}
P(C_{\rm max}) = \frac{1}{2\rho+3}exp\{[\frac{\phi+\sqrt{\phi^2-8C_{\rm
        max}(6\rho+1)}}{2C_{\rm
      max}}-2(\rho+1)]ln(\frac{2\rho+2}{2\rho+3})\}.
\end{array}
\end{equation*}

\textbf{Theorem 5.} [34] We let $\{x_{t}\}_{t=0,1,...,N}$ be a bi-finite
sequence of $i.i.d.$ random variables extracted from a continuous
probability density $f(x)$. Then the probability $P_{\rho}(n)$ that two
data separated by $n$ intermediate data are two connected nodes in the
LPHVG$(\rho)$ is
\begin{equation*}
\begin{array}{l}
P_{\rho}(n) = \frac{2\rho(\rho+1)+2}{n(n+1)},\rho=0,1,2,...
\end{array}
\end{equation*}

Theorem 4 shows the distribution characteristics of the minimum
clustering coefficient and the maximum clustering coefficient of the
nodes in LPHVG$(\rho)$. Theorem 5 indicates that the limited penetrable
visibility probability $P_{\rho}(n)=\frac{2\rho(\rho+1)+2}{n(n+1)}$
introduces shortcuts in the LPHVG$(\rho)$. With these shortcuts the
limited penetrable horizontal visibility graph reveals the presence of
small-world phenomena \cite{34}.

\textbf{Theorem 6.} We let $X(t)$ be a bi-infinite sequence of $i.i.d.$
of random variable $X$ with a probability density $f(x)$. Then both the
in and out degree distribution of its associated DLPHVG$(\rho)$ is
\begin{equation*}\label{eq1}
P_{\rm in}(k) = P_{\rm out}(k) =
\begin{cases}
\frac{1}{\rho+2}(\frac{\rho+1}{\rho+2})^{k-(\rho+1)},k\geq \rho+1.\\
0, otherwise.
\end{cases}
\end{equation*}

\textbf{Proof.} Examining the out-distribution $P_{\rm out}(k)$ we let
$x$ be an arbitrary datum of the $i.i.d.$ random time series, and
$x_{br}\geq x$ the probability that its limited penetrable horizontal
visibility is interrupted by one bounding datum on its right. There are
$\rho$ penetrable data $x_{p1},x_{p2},...,x_{p\rho} \geq x$ between $x$
and the bounding data $x_{br}$. These $\rho+1$ data are independent of
$f(x)$. Then
\begin{equation}\label{eq1}
\begin{array}{l}
D\Phi_{\rm
  out}^{\rho+1}=\int_{-\infty}^{\infty}\int_{x}^{\infty}...
\int_{x}^{\infty}\int_{x}^{\infty}f(x)f(x_{p1})...f(x_{p\rho})
f(x_{br})dx_{br}dx_{p\rho}...dx_{p1}dx\\
=\int_{-\infty}^{\infty}f(x)[1-F(x)]^{\rho+1}dx=\frac{1}{\rho+2}.
\end{array}
\end{equation}
The probability $P_{\rm out}(k)$ that datum $x$ penetrates no more than
$\rho$ time seeing $k$ data is
\begin{equation}\label{eq1}
\begin{array}{l}
P_{\rm out}(k) = D\Phi_{\rm out}(k)D\Phi_{\rm out}^{\rho+1}=
\frac{1}{\rho+2}D\Phi_{\rm out}(k),
\end{array}
\end{equation}
where $D\Phi(k)$ is the probability that $x$ penetrates no more than
$\rho$ time to the right seeing at least $k$ data. Then $D\Phi(k)$ can
be recurrently calculated
\begin{equation}\label{eq1}
\begin{array}{l}
D\Phi(k)=D\Phi(k-1)(1-D\Phi_{\rm
  out}^{\rho+1})=\frac{\rho+1}{\rho+2}D\Phi(k-1)
=(\frac{\rho+1}{\rho+2})^{k-(\rho+1)}D\Phi(\rho+1),
\end{array}
\end{equation}
from which, with $D\Phi(\rho+1)=1$, we deduce
\begin{equation}\label{eq1}
\begin{array}{l}
D\Phi(k)=(\frac{\rho+1}{\rho+2})^{k-(\rho+1)}.
\end{array}
\end{equation}
Thus we finally obtain
\begin{equation}\label{eq1}
\begin{array}{l}
P_{\rm out}(k) = D\Phi_{\rm out}(k)D\Phi_{\rm out}^{\rho+1}=
\begin{cases}
\frac{1}{\rho+2}(\frac{\rho+1}{\rho+2})^{k-(\rho+1)}.\\
0,otherwise.
\end{cases}
\end{array}
\end{equation}

To further check the accuracy of Eq.~(21), we perform several numerical
simulations. We generate random series of 3000 data from uniform,
gaussian, and power law distributions and their associated
DLPHVG$(\rho)$. Fig.~6 show plots of the degree distributions with
penetrable distances $\rho = 0$, 1, and 2. Circles indicate $P_{\rm
  in}(k)$, diamonds $P_{\rm out}(k)$), and the solid line the
theoretical results of Eq.~(21). We find that the theoretical results
agree with the numerics, placing aside finite size effects. As in the
degree distribution of LPHVG$(\rho)$ \cite{34}, the deviations between
the tails of the in and out degree distributions of DLPHVG$(\rho)$
associated with $i.i.d.$ random series are caused solely by finite size
effects.

\begin{figure}[H]
\centering \scalebox{0.5}[0.5]{\includegraphics{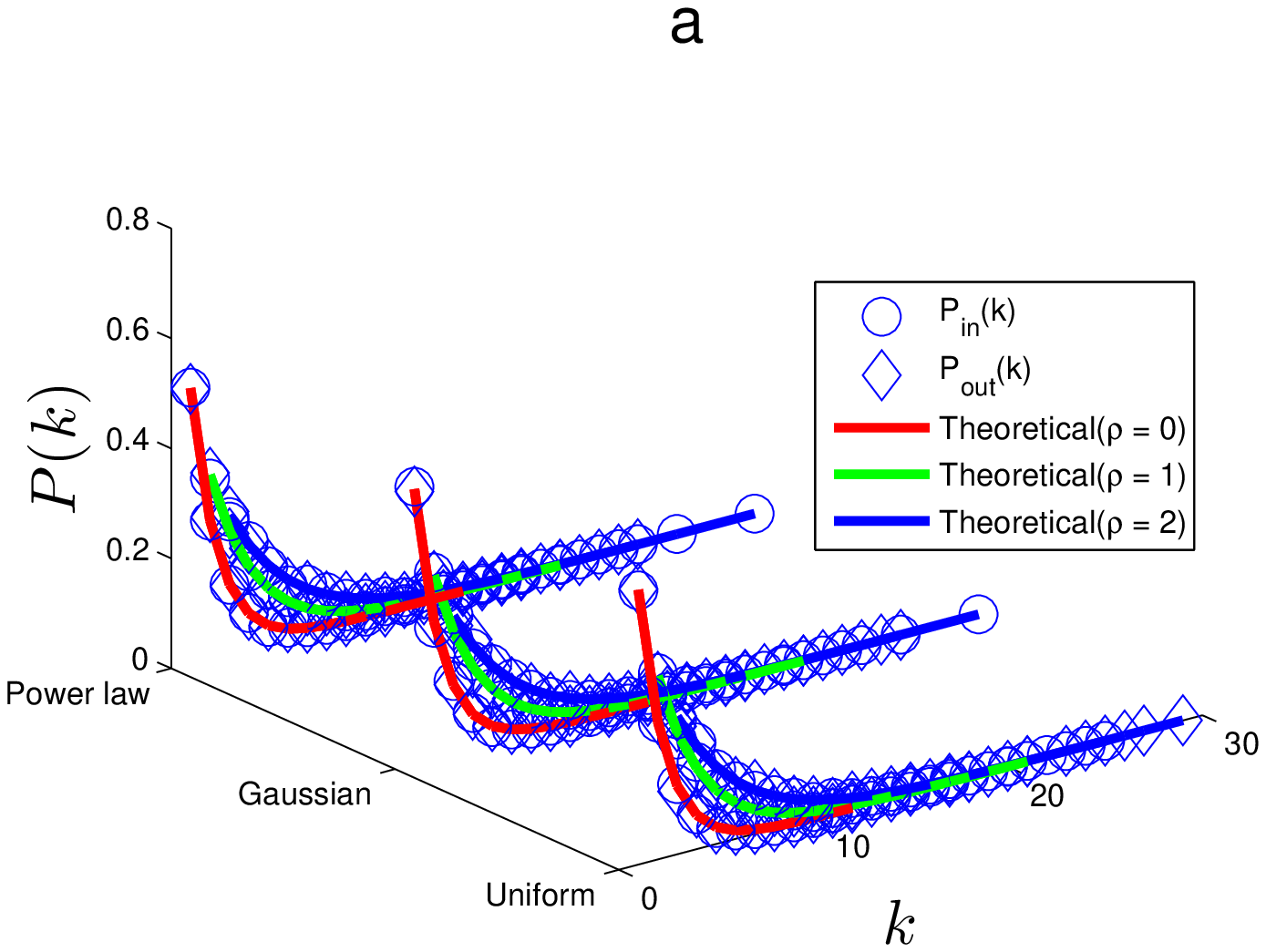}}
\scalebox{0.5}[0.5]{\includegraphics{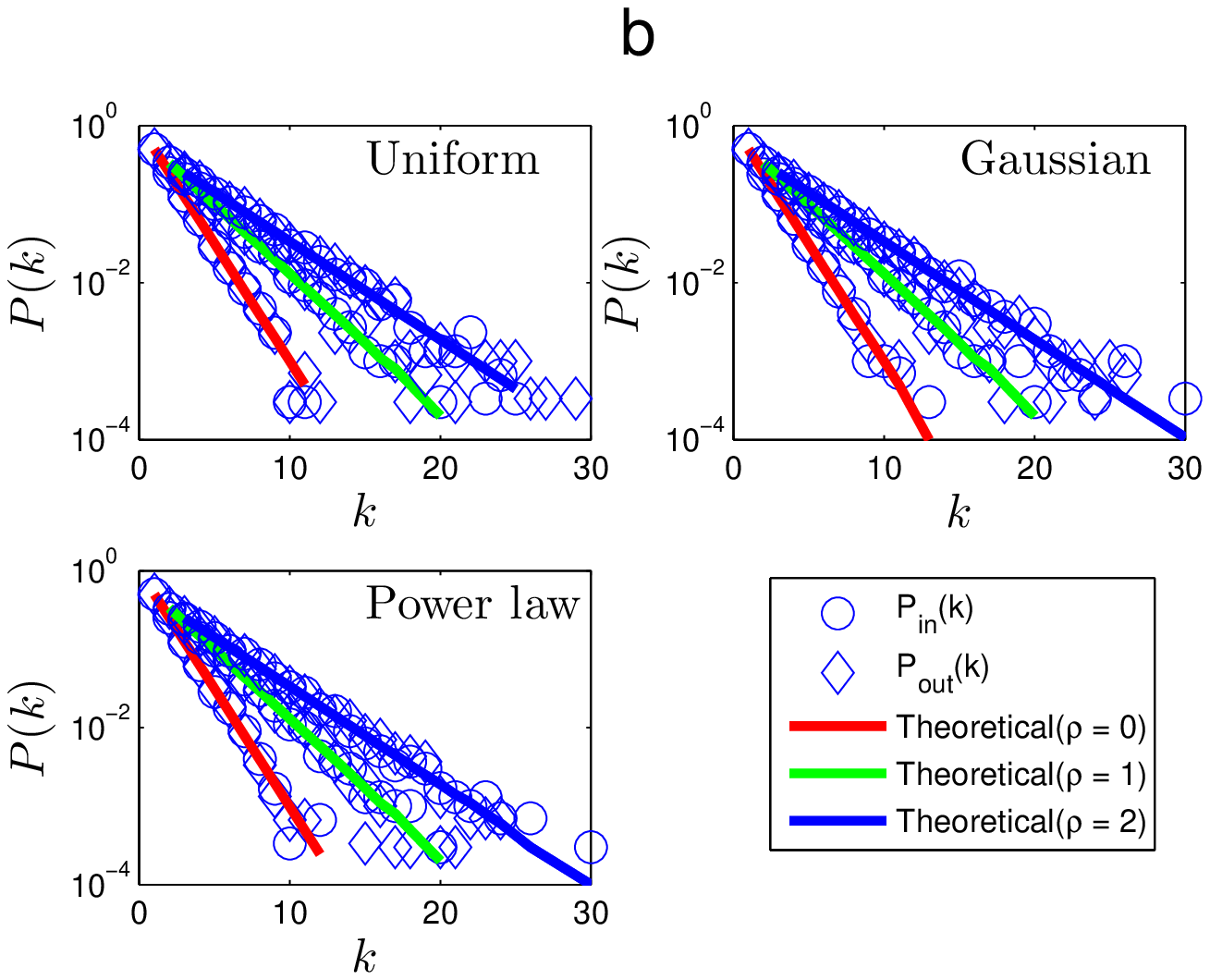}}
\end{figure}

\begin{figure}[H]
\caption{(a) Plot of the in and out degree distribution of the
    resulting graphs, (b) semi-log plot of the in and out degree
    distribution of the resulting graphs.}
\end{figure}

\textbf{Theorem 7.} We let $\textbf{X}_{N \times N}$ be a matrix with
entries $x_{ij}=\xi$, where $\xi$ is a random variable sampled from a
distribution $f(x)$. Then when $n>0$ and in the limited $N\rightarrow
\infty$, the degree distribution of the associated ILPHVG$_{n}(\rho)$
converges to
\begin{equation*}\label{eq1}
\begin{array}{l}
P(k)=
\begin{cases}
\frac{1}{[n(\rho+1)+1]}[\frac{n(\rho+1)}{n(\rho+1)+1}]^{k-n(\rho+1)},
k\geq n(\rho+1).\\
0,otherwise.
\end{cases}
\end{array}
\end{equation*}

\textbf{Proof.} To derive general results, we consider the two special
cases $n=4$ and $n=8$.

In the case $n = 4$, we let $x$ be an arbitrary datum in $\textbf{R}_{N
  \times N}$ where the probability of its image limited penetrable
horizontal visibility is interrupted by four bounding datum, i.e.,
$x_{br}$ on its right, $x_{bu}$ above it, $x_{bl}$ on its left, and
$x_{bb}$ below it. There are $4\rho$ penetrable data
$x_{pr1},...,x_{pr\rho},x_{pu1},...,x_{pu\rho},x_{pl1},...,x_{pl\rho},x_{pb1}...,x_{pb\rho}$
between $x$ and the four bounding data. These $4\rho+4$ data are
independent of $f(x)$. Then
\begin{equation}\label{eq1}
\begin{array}{l}
I\Phi[4(\rho+1)]=\int_{-\infty}^{\infty}\int_{x}^{\infty}...\int_{x}^{\infty}\int_{x}^{\infty}...\int_{x}^{\infty}\int_{x}^{\infty}...\int_{x}^{\infty}\int_{x}^{\infty}...\int_{x}^{\infty}\int_{x}^{\infty}\int_{x}^{\infty}\int_{x}^{\infty}\int_{x}^{\infty}f(x)f(x_{pr1})...f(x_{pr\rho})f(x_{pu1})...\\
...f(x_{pu\rho})f(x_{pl1})...f(x_{pl\rho})f(x_{pb1})...f(x_{pb\rho})f(x_{br})f(x_{bu})f(x_{bl})f(x_{bb})dx_{bb}dx_{bl}dx_{bu}dx_{br}dx_{pb\rho}...\\
...dx_{pb1}dx_{pl\rho}...dx_{pl1}dx_{pu\rho}...dx_{pu1}dx_{pr\rho}...dx_{pr1}dx\\
=\int_{-\infty}^{\infty}f(x)[1-F(x)]^{4\rho+4}dx = \frac{1}{4\rho+5}.
\end{array}
\end{equation}
The probability that the node $x$ has a penetrable visibility of exactly
$k$ nodes is
\begin{equation}\label{eq1}
\begin{array}{l}
P(k)=
\{1-I\Phi[4(\rho+1)]\}^{k-4(\rho+1)}I\Phi[4(\rho+1)]
=\frac{1}{4\rho+5}(\frac{4\rho+4}{4\rho+5})^{k-4(\rho+1)},
k\geq 4(\rho+1).
\end{array}
\end{equation}
Similarly, when $n=8$ from Eq.~(22), then
\begin{equation}\label{eq1}
\begin{array}{l}
I\Phi[8(\rho+1)]=\int_{-\infty}^{\infty}f(x)[1-F(x)]^{8\rho+8}dx = \frac{1}{8\rho+9}.
\end{array}
\end{equation}
Here the probability that node $x$ has a penetrable visibility of
exactly $k$ nodes is
\begin{equation}\label{eq1}
\begin{array}{l}
P(k)=
\{1-I\Phi[8(\rho+1)]\}^{k-8(\rho+1)}I\Phi[8(\rho+1)]
=\frac{1}{8\rho+9}(\frac{8\rho+8}{8\rho+9})^{k-8(\rho+1)},
k\geq 8(\rho+1).
\end{array}
\end{equation}
From Eqs.~(23) and (25) we deduce a generic $n$ that yields
\begin{equation}\label{eq1}
\begin{array}{l}
P(k)=
\{1-I\Phi[n(\rho+1)]\}^{k-n(\rho+1)}I\Phi[n(\rho+1)]
=
\begin{cases}
\frac{1}{[n(\rho+1)+1]}[\frac{n(\rho+1)}{n(\rho+1)+1}]^{k-n(\rho+1)},
k\geq n(\rho+1).\\
0,otherwise.
\end{cases}
\end{array}
\end{equation}

Note that when $n=2$ this result reduces to that in Eq.~(10). To check
the accuracy of Eq.~(26), we estimate the degree distribution of
ILPHVG$_{n}(\rho)$ associated with $N \times N$ random matrices whose
entries are $i.i.d.$ uniform random variables $U[0,1]$. To illustrate
the finite size effects, we also define the cutoff value. When $k>k_{0}$
all the degree distributions of the numerical results are smaller than
the theoretical result in Eq.~(26), and $k_{0}$ is the cutoff
value. Figs.~7(a) and 7(c) show semi-log plots of the finite size
degree distributions of ILPHVG$_{4}(\rho)$ and ILPHVG$_{8}(\rho)$ with
$N = 200$. Note that the distributions agree with Eq.~(26) when $k \leq
k_{0}$. To assess the convergence speed of Eq.~(26) for finite $N$, we
estimate the cutoff value $k_{0}$ under different finite $N$ sizes [see
  Figs.~7(b) and 7(d)]. Note that the location of the cutoff value
$k_{0}$ scales logarithmically with the system size $N$, i.e., finite
size effects only affect the tail of the distribution, which quickly
converges logarithmically with $N$.

\begin{figure}[H]
\centering \scalebox{0.42}[0.42]{\includegraphics{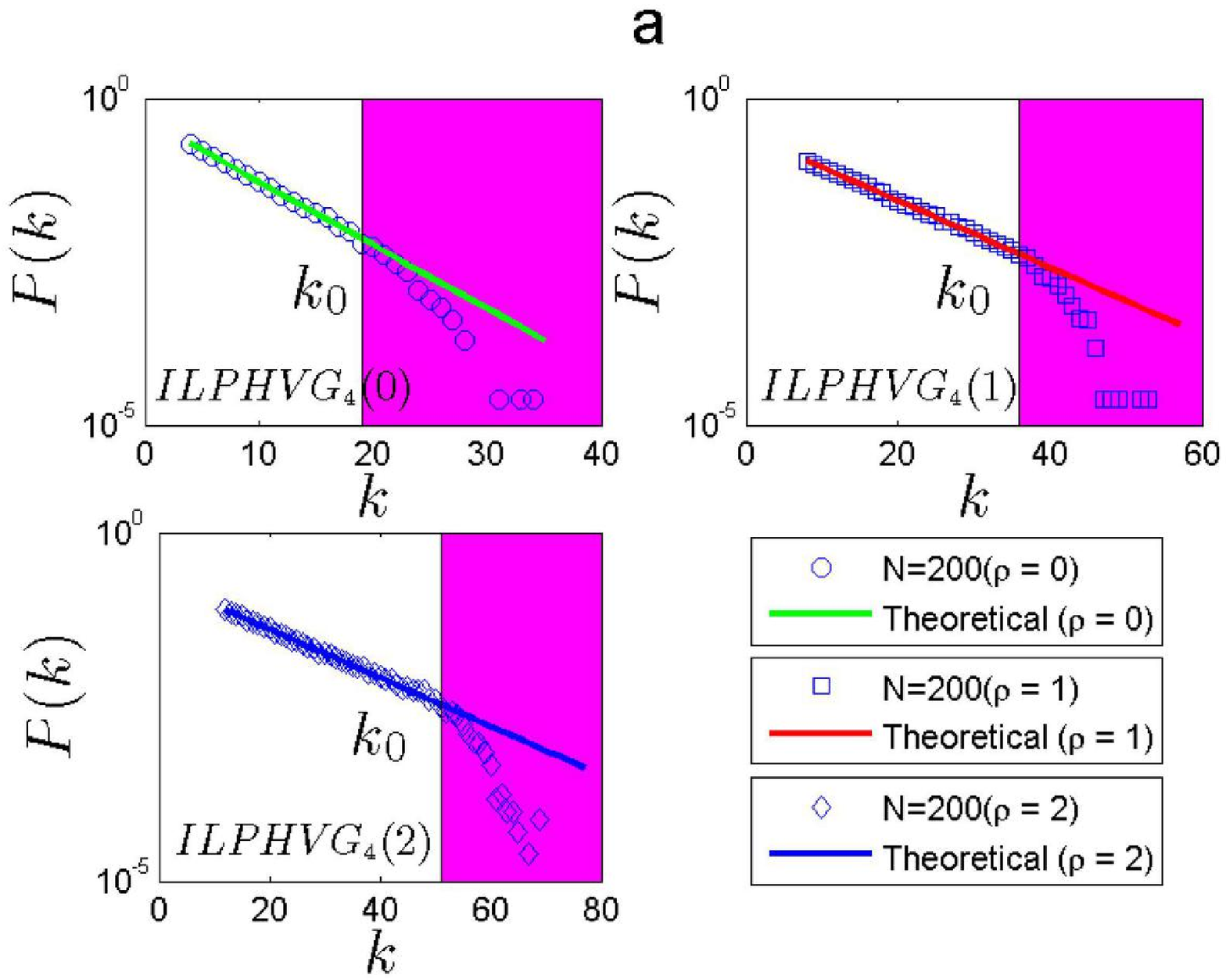}}
\scalebox{0.6}[0.6]{\includegraphics{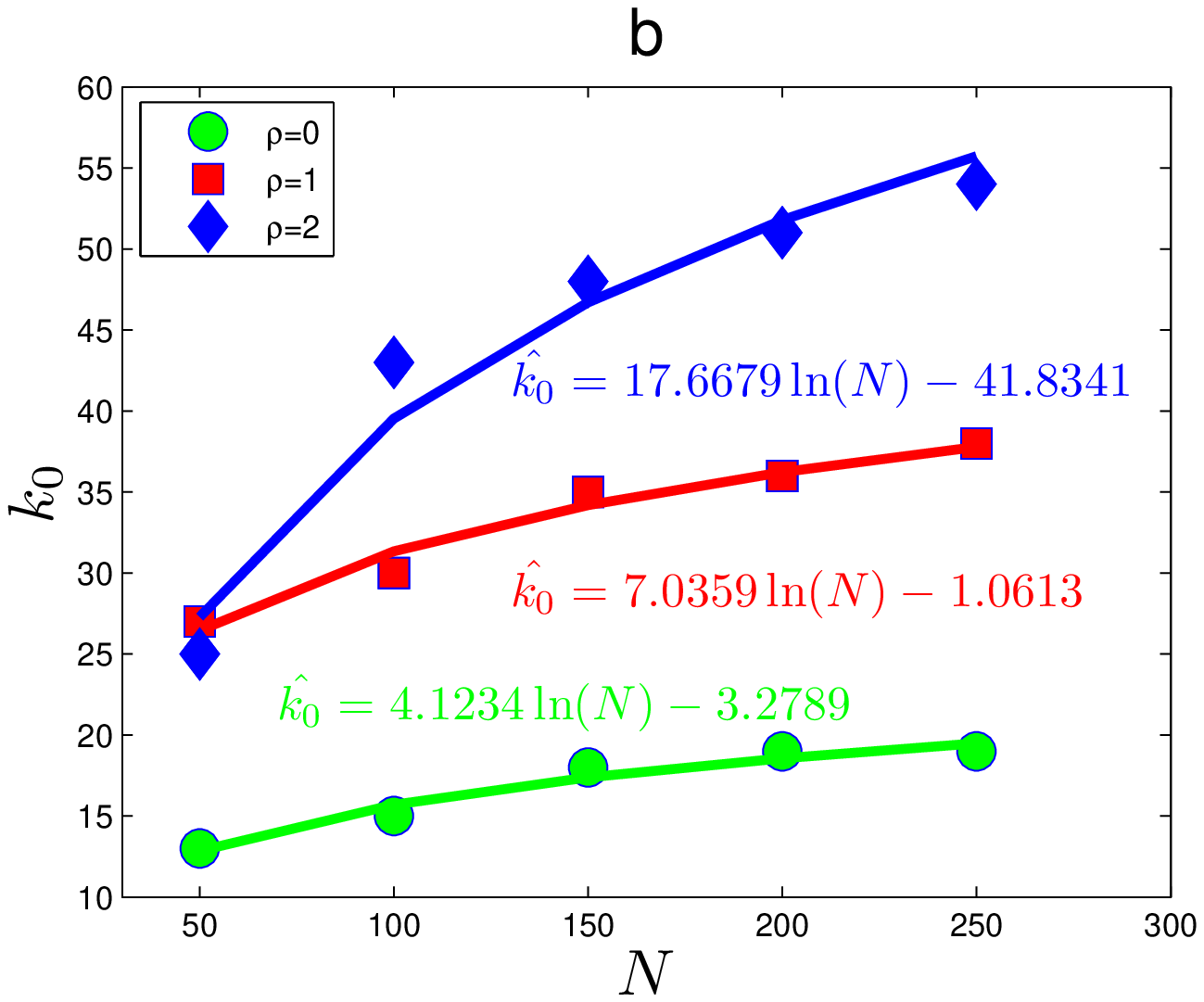}}
\scalebox{0.42}[0.42]{\includegraphics{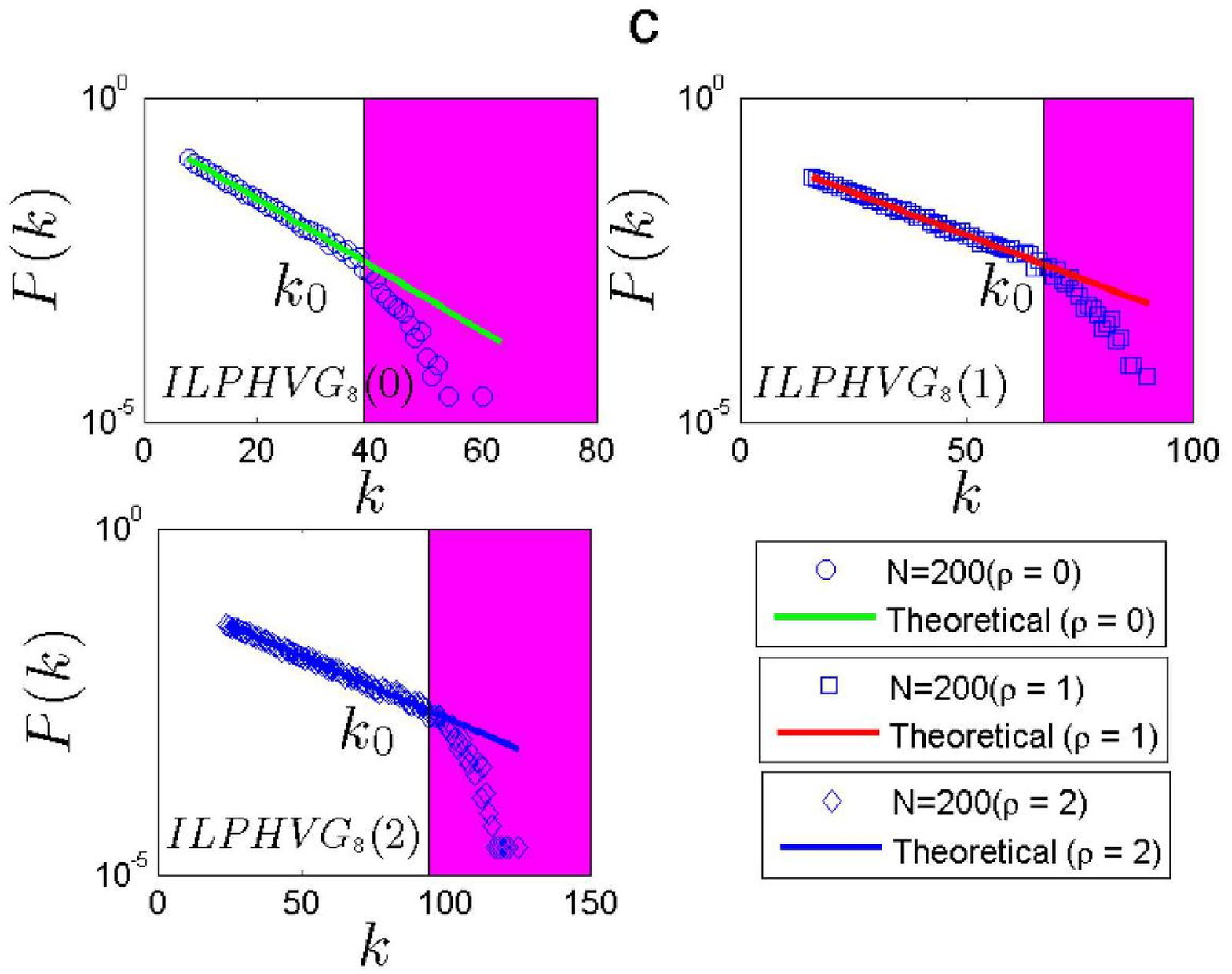}}
\scalebox{0.6}[0.6]{\includegraphics{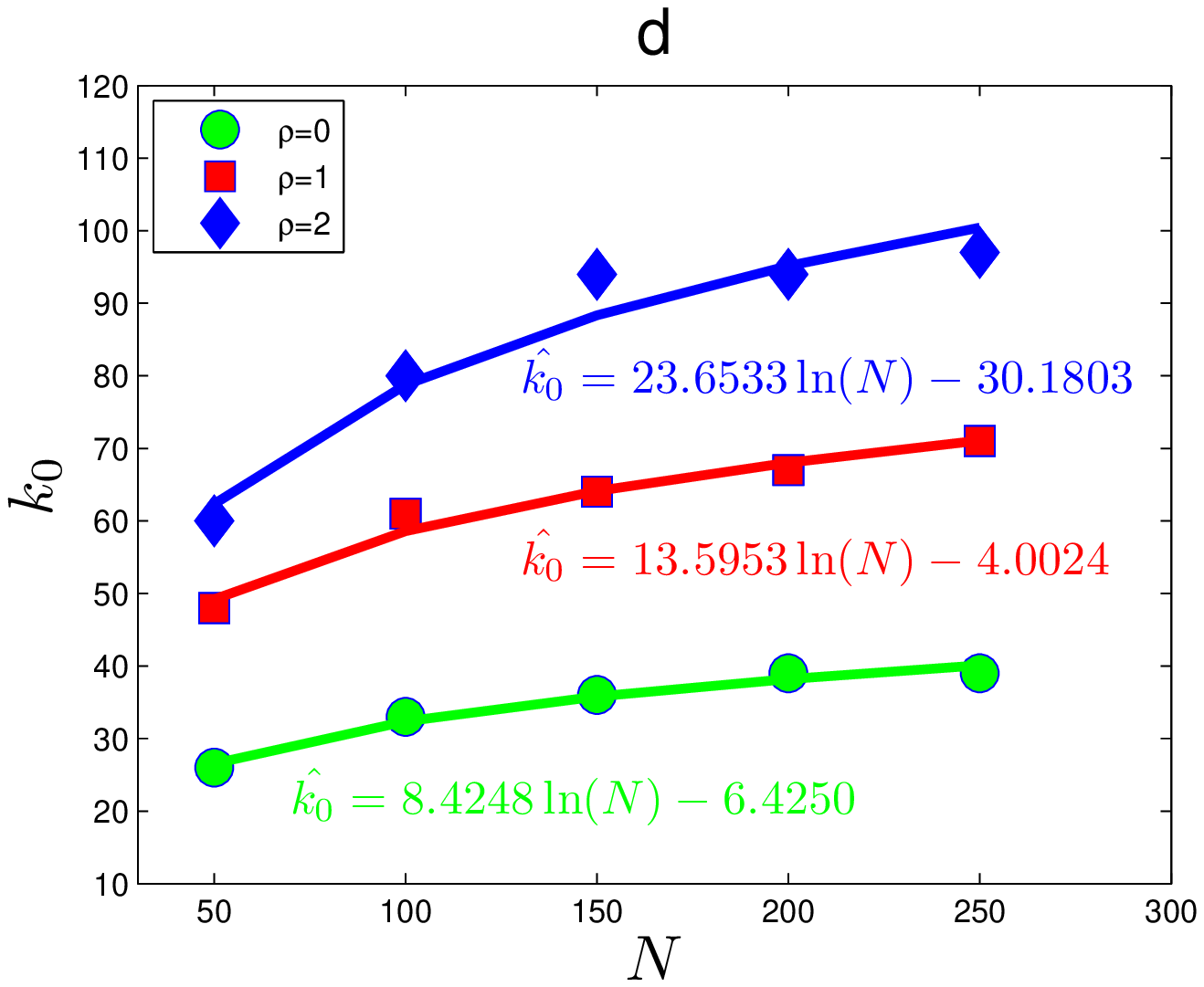}}
\end{figure}
\begin{figure}[H]
\caption{( a ) Semi-log plot of the degree distribution of
    ILPHVG$_{4}(\rho)$ associated to $N \times N$ random matrices. The
    solid line is the theoretical value of $P(k)$ given by Eq. (26). In
    every case we find excellent agreement with Eq. (26) for $k \leq
    k_{0}$, where $k_{0}$ is a cutoff value that denotes the onset of
    finite size effects. ( b ) plot of the cutoff $k_{0}$ as a function
    of different size $N$, suggesting a logarithmic
    scaling. ( c ) semi-log plot of the degree distribution of
    ILPHVG$_{8}(\rho)$ associated to $N \times N$ random matrices. The
    solid line is the theoretical value of $P(k)$ given by Eq. (26). In
    every case we find excellent agreement with Eq. (26) for $k \leq
    k_{0}$. ( d ) plot of the cutoff $k_{0}$ as a function of different size $N$, suggesting a logarithmic scaling.}
\end{figure}

\section{Application of DLPHVG$(\rho)$ and ILPHVG$_{n}(\rho)$)}

We use the analytical results of LPHVG$(\rho)$ to distinguish between
random and chaotic signals [34], and we describe the global evolution of
crude oil futures. We also describe applications of DLPHVG$(\rho)$
and ILPHVG$_{n}(\rho)$.

\textbf{Measure real-valued time series irreversibility by
  DLPHVG$(\rho)$}. Time series irreversibility is an important topic in
basic and applied science \cite{35}. Over the past decade several
methods of measuring time irreversibility have been proposed
\cite{36,37,38}. A recent proposal uses the directed horizontal
visibility algorithm \cite{39}. Here the Kullback-Leibler divergence
(KLD) between the out- and in-degree distributions is defined
\begin{equation}\label{eq1}
\begin{array}{l}
D[P_{\rm out}(k)||P_{\rm in}(k)] = \sum\limits_{k}P_{\rm
  out}(k)log\frac{P_{\rm out}(k)}{P_{\rm in}(k)}.
\end{array}
\end{equation}

Eq.~27 measures the irreversibility of real-value stationary
stochastic series, and we here explore the applicability of
DLPHVG$(\rho)$. We first select an appropriate parameter $\rho$, map a
time series to a directed limited penetrable horizontal visibility
graph, and then use Eq.~27 to estimate the degree of irreversibility of
the series. Using Theorem 6 and Eq.~27 we find that the KLD between the
in- and out-degree distributions associated with an $i.i.d.$ random
infinite series is equal to zero. Using our analysis of finite size
effects, we infer that the KLD between the in- and out-degree
distributions associated with an $i.i.d.$ random finite series of size
$N$ tends asymptotically to zero. We set $\rho = 0$, 1, and 2, and
calculate the numerical value of the KLD of the random series of 3000
data from uniform, Gaussian, and power-law distributions (see the upper
section of Table~1). All numerical values of KLD are approximately 0,
which suggests that the $i.i.d.$ time series is reversible.

We next examine the chaotic logistic $(\mu=4)$ and H\'{e}non $(a=1.4,
b=0.3)$ map series. Figures~8(a) and 8(b) show plots of the in- and
out-degree distributions of DLPHVG($\rho$), $\rho=0,1,2$ associated with
the Logistic map at $\mu=4$ and the H\'{e}non map at $a=1.4$ and $b=0.3$
of 3000 data points. Note that in each case there is a clear distinction
between the in- and out-degree distributions, and this differs from the
$i.i.d.$ series case [see Fig.~6(b)]. We calculate the values of KLD for
each case (bottom section of Table~1). We find that the values of KLD
are positive and much larger than those of the $i.i.d.$
series. Figs.~8(c) and 8(d) show a finite size analysis of the chaotic
maps. Note that the KLD values associated with the chaos maps converges
with series size $N$ to a asymptotical nonzero value, which indicates
that chaos maps are irreversible.

\begin{table}\caption{\emph{Values of the irreversibility measure
      associated to the degree distribution $D[P_{\rm out}(k)||P_{\rm
          in}(k)]$ for DLPHVG$(\rho)$ associated to series of 3000 data
      generated from reversible and irreversible processes}}
\centering
\label{tab:your-surname1}
\begin{tabular}{llll}
\hline
Series description & $\rho = 0$ & $\rho = 1$ & $\rho = 2$ \\
\hline
Uniform distribution $f(x) = U[0,1]$        & 0.000950    &  0.007106  &  0.007269\\
Gaussian distribution                       &  0.002633   & 0.007106   &  0.005507\\
Power law distribution $f(x) \sim x^{-2}$   & 0.000226    & 0.004257  &  0.005267\\

\hline
Logistic map ($\mu = 4$)                    &  0.342985  & 0.090773 & 0.081985\\
H\'{e}non map ($a = 1.4,b = 0.3$)           &  0.158358      &  0.125637    & 0.140270    \\

\hline

\end{tabular}
\begin{flushleft}
\end{flushleft}
\end{table}

Thus by selecting an appropriate parameter for $\rho$, the $D[P_{\rm
    out}(k)||P_{\rm in}(k)]$ of DLPHVG$(\rho)$ captures the
irreversibility of the time series.

\begin{figure}[H]
\centering \scalebox{0.6}[0.6]{\includegraphics{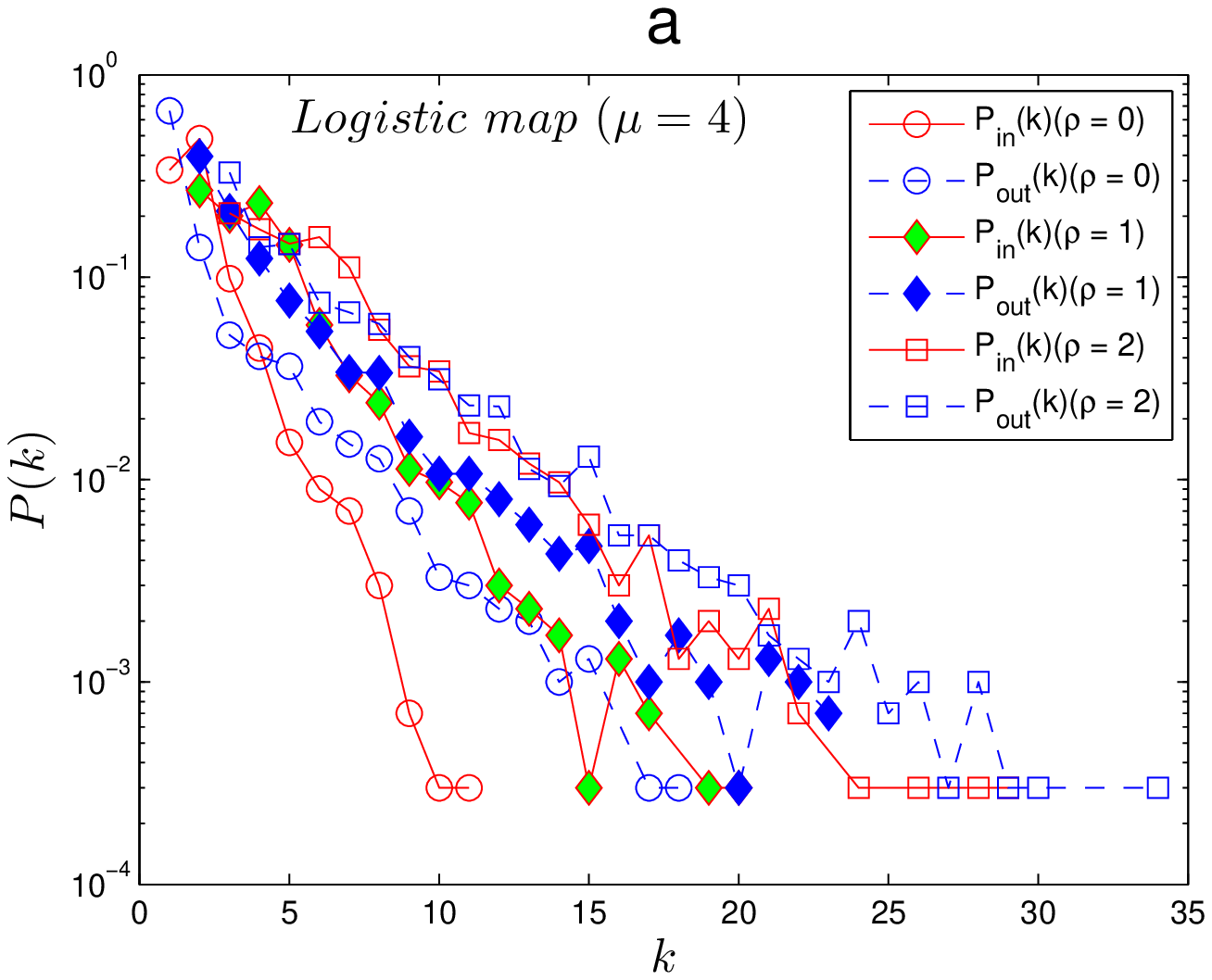}}
\scalebox{0.6}[0.6]{\includegraphics{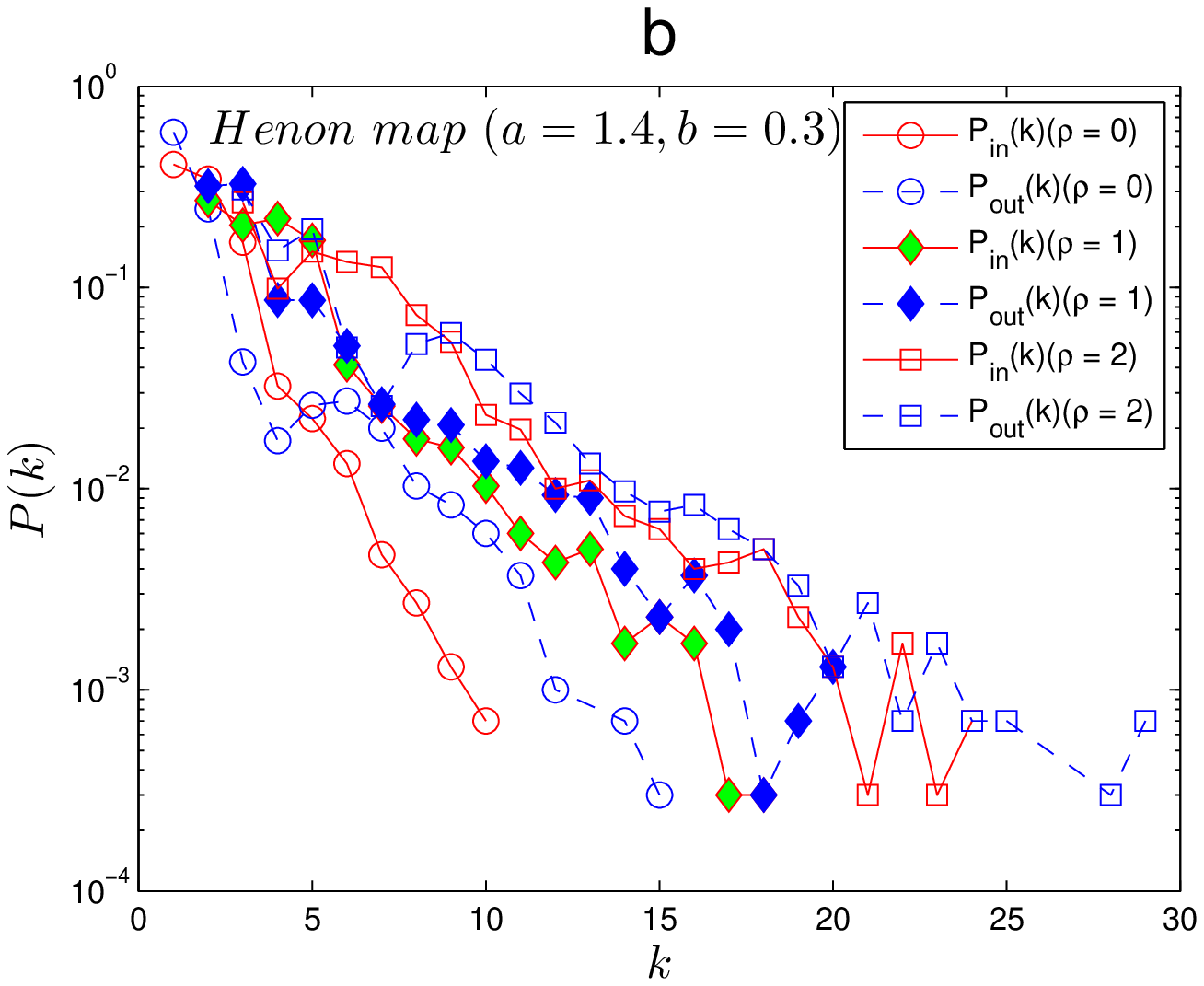}}
\scalebox{0.6}[0.6]{\includegraphics{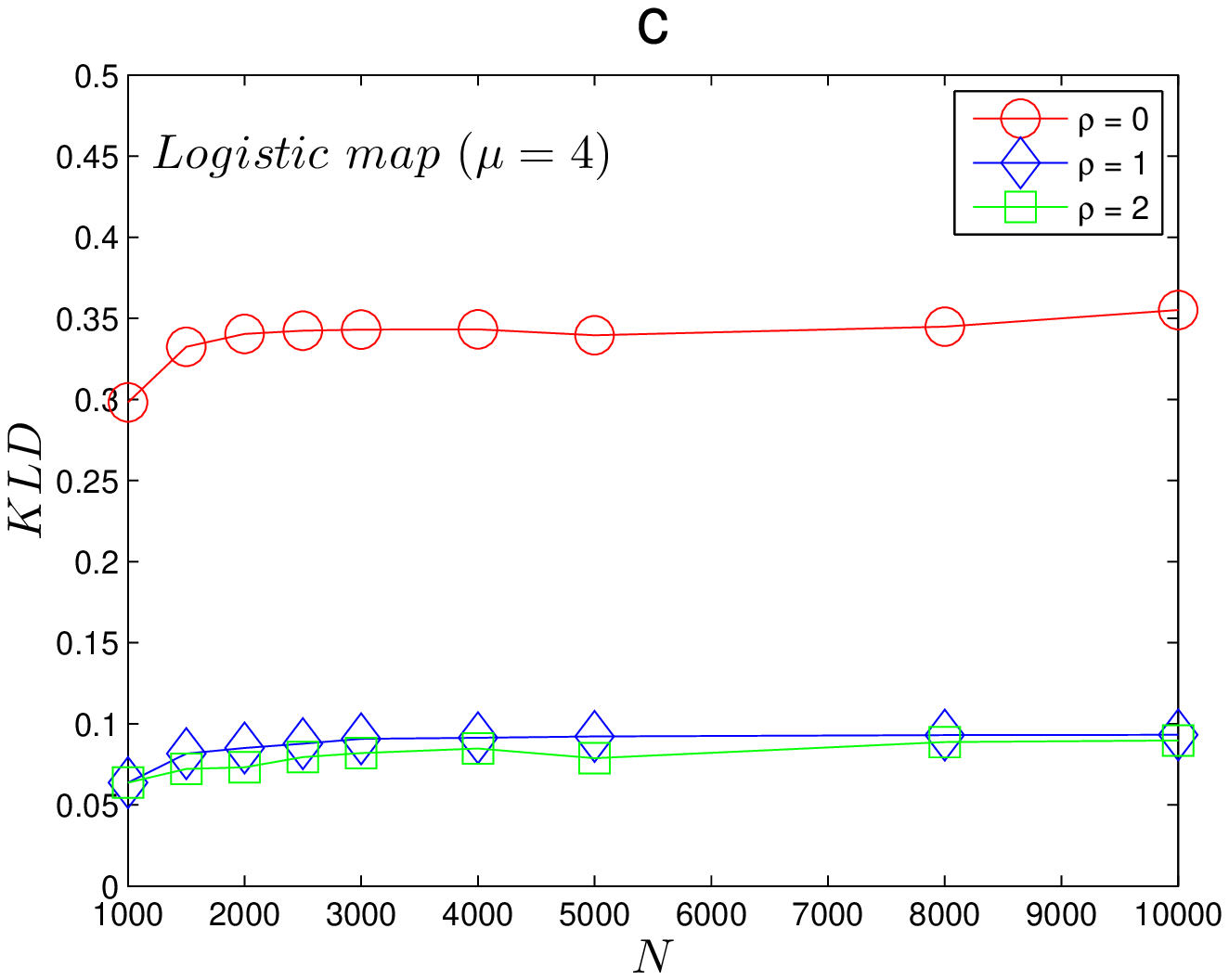}}
\scalebox{0.6}[0.6]{\includegraphics{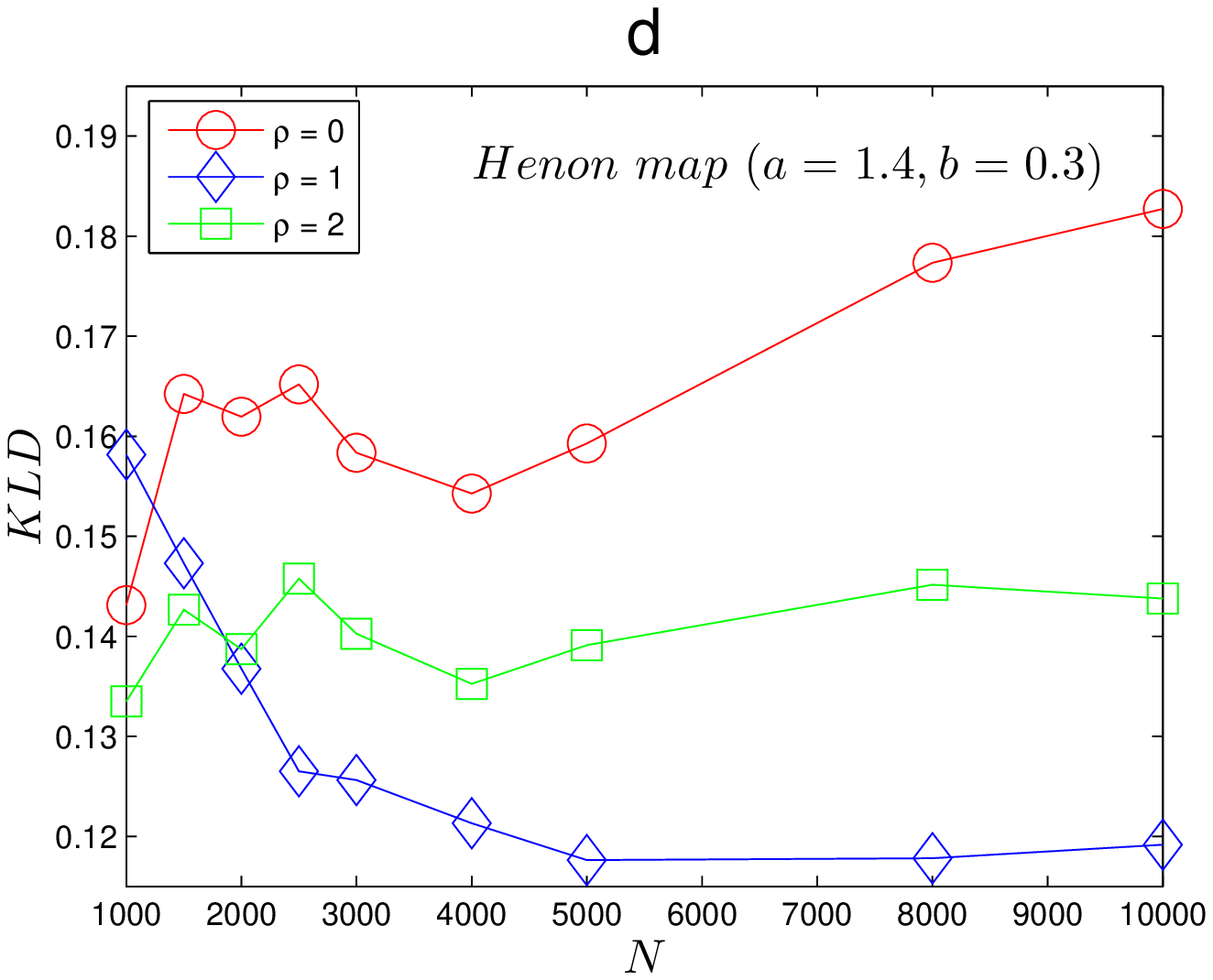}}
\end{figure}
\begin{figure}[H]
\caption{(a) Plot of the in- and out- degree distributions of
    DLPHVG($\rho$), $\rho=0,1,2$ associated to the Logistic map at
    $(\mu=4)$ of 3000 data points, (b) plot of the in- and out- degree
    distributions of DLPHVG($\rho$), $\rho=0,1,2$ associated to the
    H\'{e}non map at $( a=1.4, b=0.3)$ of 3000 data points, which
    different from the uncorrelated cases (see Fig. 6 (b)). (c) the values
    of $KLD$ associated to the Logistic map with series size $N$, (d)
    the values of $KLD$ associated to the H\'{e}non map with series size
    $N$, which converge to a asymptotical nonzero value.}
\end{figure}

\textbf{Discriminating between and chaos using
  ILPHVG$_{n}(\rho)$}. Although chaotic processes display an irregular
and unpredictable behavior that is frequently perceived to be random,
chaos is a deterministic process that often hides patterns that can be
extracted using appropriate techniques. In recent decades research
efforts to distinguish between noise and chaos have been widespread
\cite{40}, and applications have been developed in all scientific
disciplines involving complex, irregular empirical signals. Lacasa et
al. \cite{33} used visibility graphs to distinguish spatiotemporal chaos
from simple randomness. We here also examine spatially extended
structures, and we explore whether ILPHVG$_{n}(\rho)$ can distinguish
distinguish spatiotemporal chaos from simple randomness.

We define $\textbf{X}(t)$ to be a two-dimensional square lattice of
$N^{2}$ diffusively-coupled chaotic maps that evolve in time [33]. In
each vertex of this coupled map lattice (CML) we allocate a fully
chaotic logistic map $x_{t+1} = Q(x_{t}),Q(x) = 4x(1-x)$, and the system
is then spatially coupled,
\begin{equation}\label{eq1}
\begin{array}{l}
\textbf{X}_{ij}(t+1) =
(1-\epsilon)Q[\textbf{X}_{ij}(t)]+\frac{\epsilon}{4}
\sum\limits_{i',j'}Q[\textbf{X}_{i'j'}(t)],
\end{array}
\end{equation}
where the sum extends to the Von Neumann neighborhood of $ij$ (four
adjacent neighbors). The update is parallel, we use periodic boundary
conditions, and the coupling strength is $\epsilon \in
[0,1]$. Fig.~9(a) shows a semi-log plot for $N=200$ of the degree
distribution of ILPHVG$_{8}(\rho),\rho =0,1,2$ associated with a
two-dimensional uncorrelated random field of uniform random variables
(stars), and a two-dimensional coupled map lattice of diffusively
coupled fully chaotic logistic maps for the coupling constants $\epsilon
= 0$ (squares), and $\epsilon = 0.1$ (diamonds). Figure~9(b) shows a
plot of the degree distribution of ILPHVG$_{8}(\rho),\rho =0,1,2$
associated with the two-dimensional coupled map lattices of diffusively
coupled fully chaotic logistic maps with a coupling constant $\epsilon =
0.7$. Eq.~(26) shows $\rho = 0$ (green line), $\rho = 1$ (red
line), and $\rho = 2$ (pink line).

\begin{figure}[H]
\centering \scalebox{0.6}[0.6]{\includegraphics{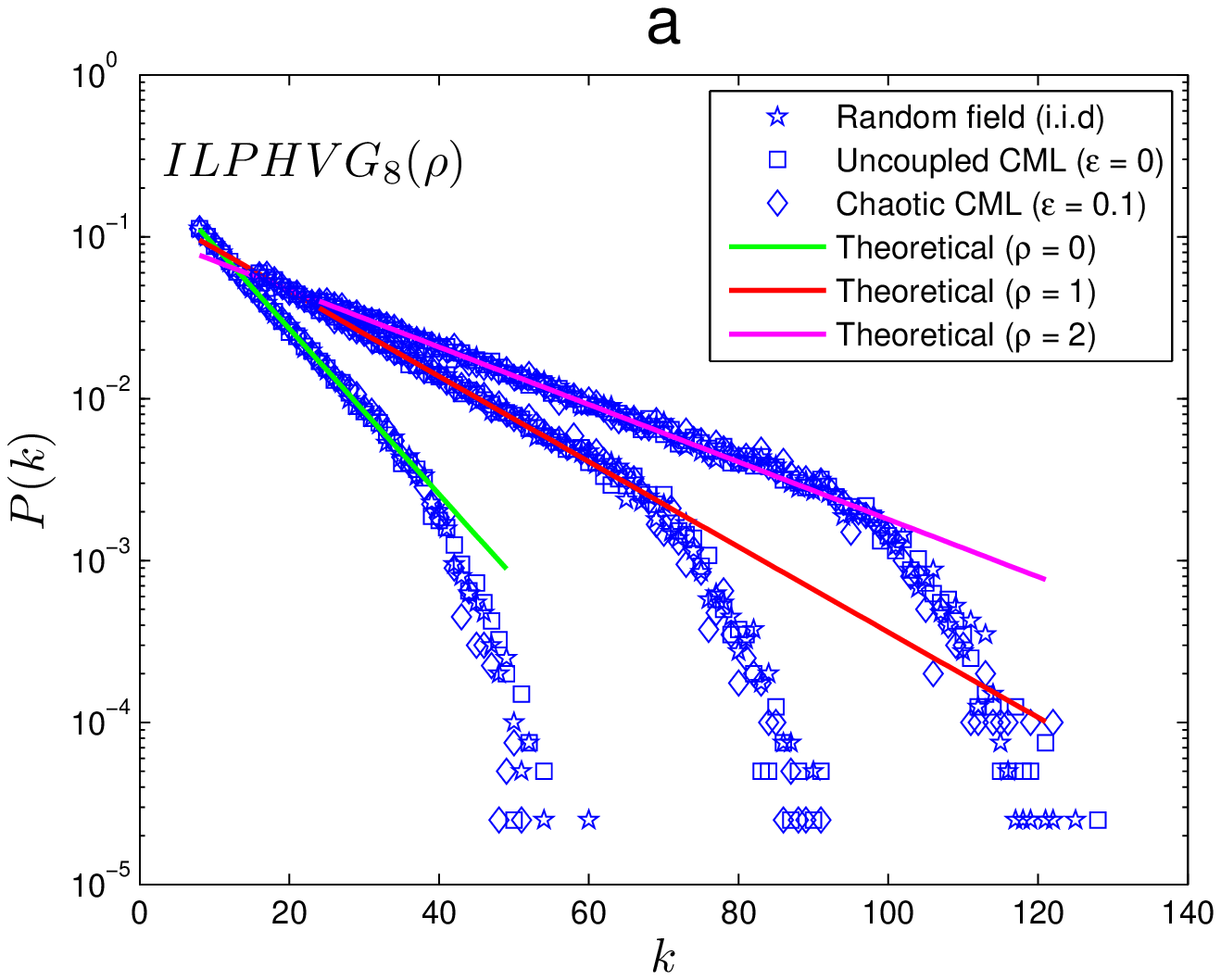}}
\scalebox{0.6}[0.6]{\includegraphics{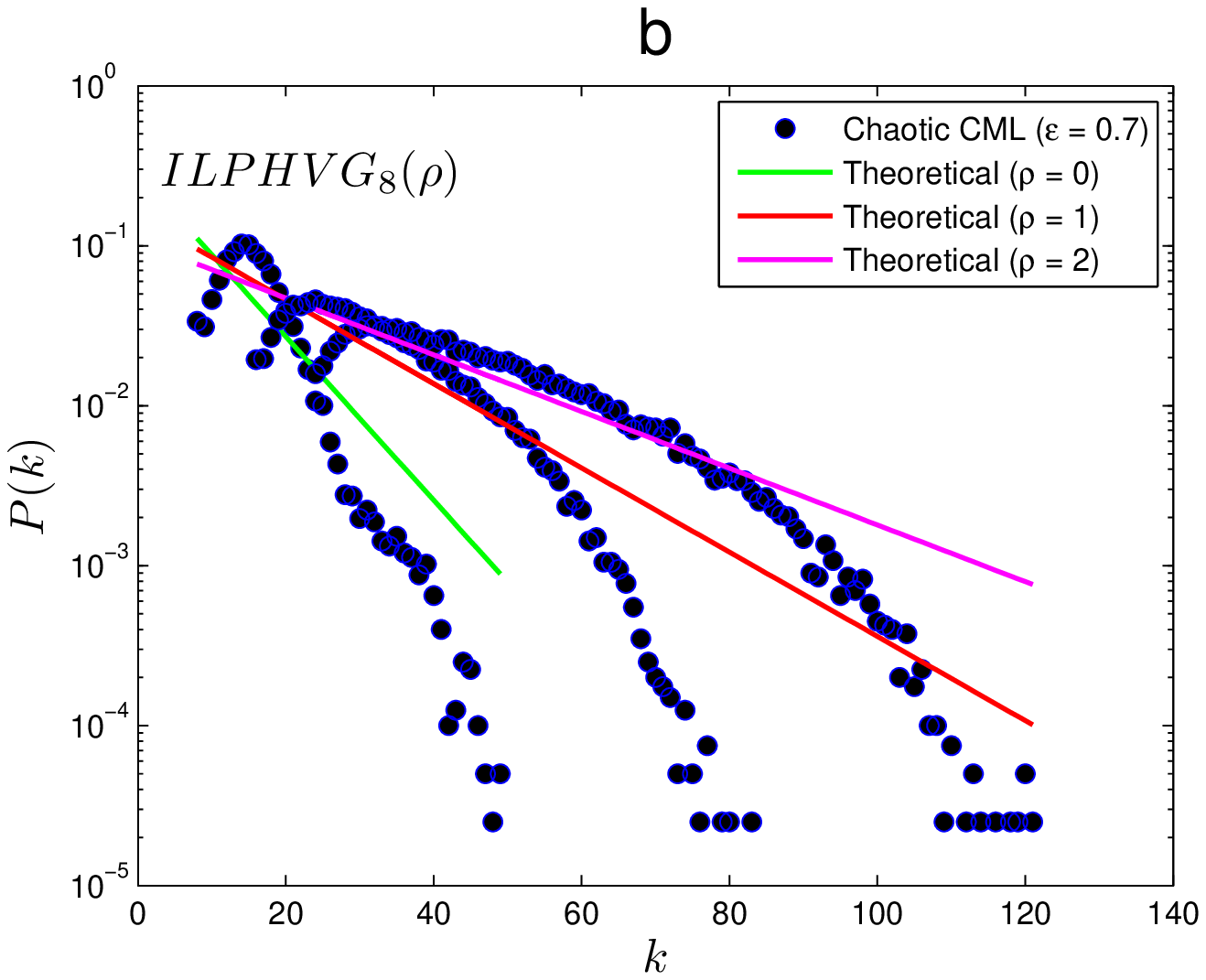}}
\scalebox{0.6}[0.6]{\includegraphics{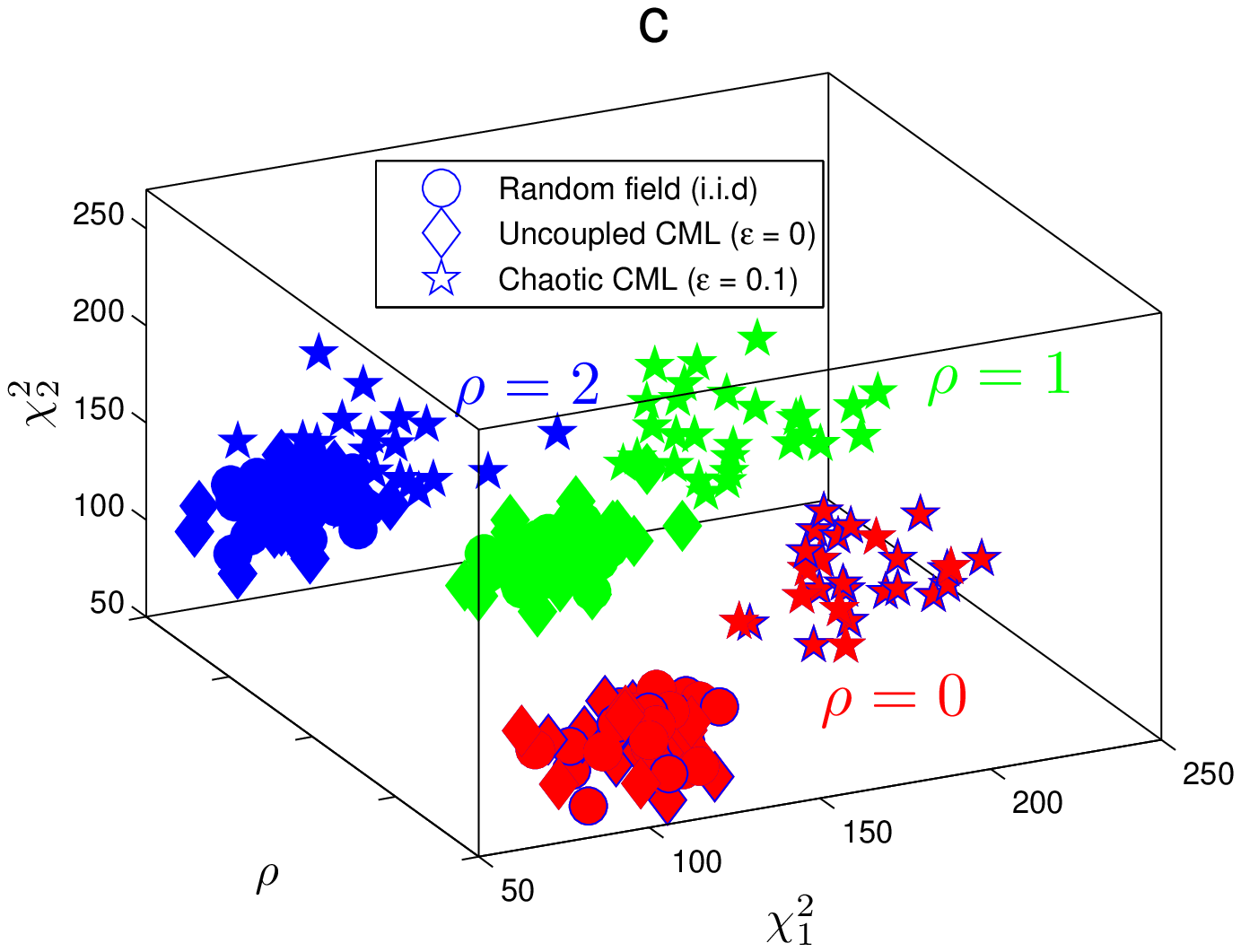}}
\scalebox{0.42}[0.42]{\includegraphics{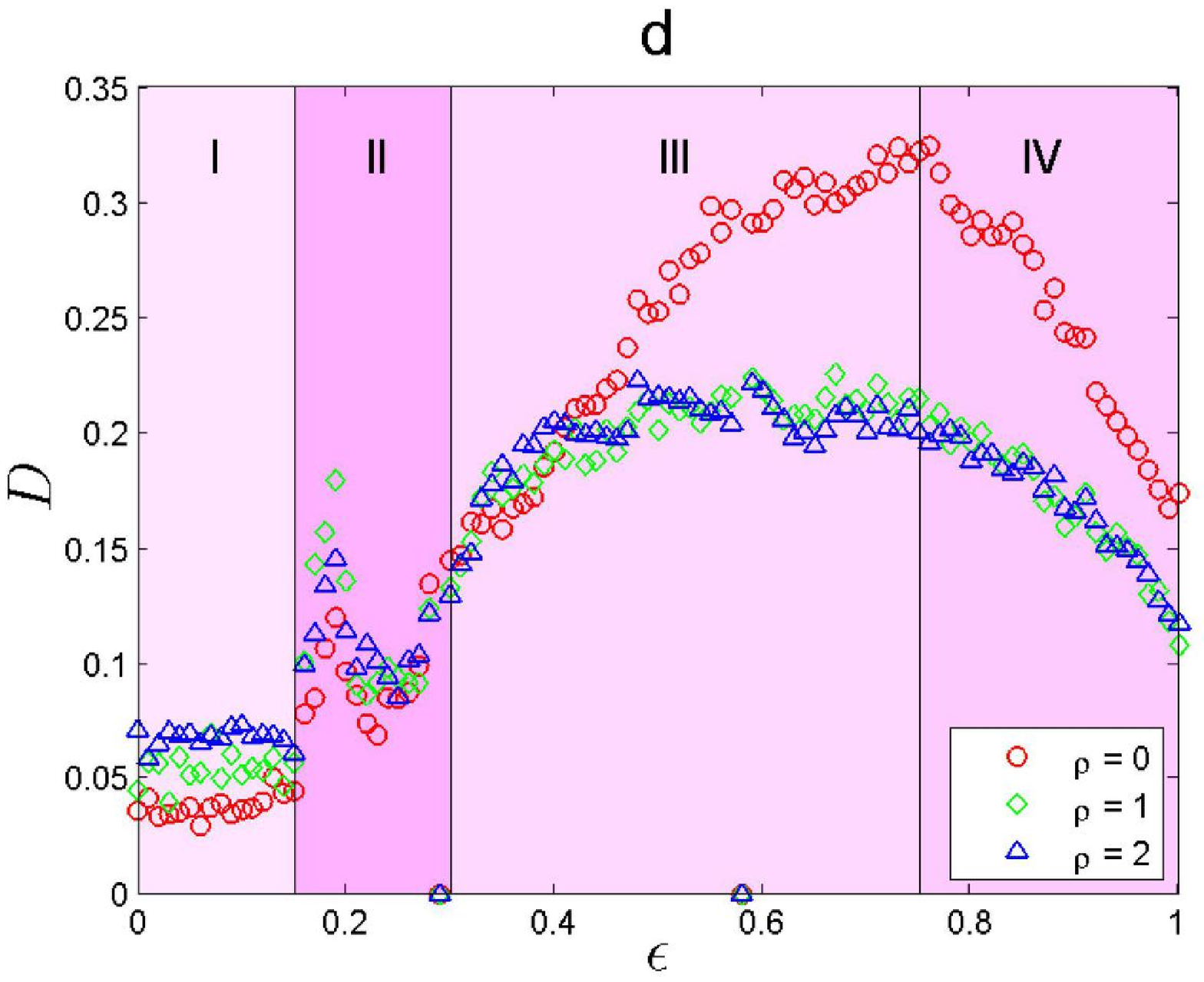}}
\end{figure}
\begin{figure}[H]
\caption{(a) Semi-log plot of the degree distribution of
    ILPHVG$_{8}(\rho), \rho=0,1,2$ associated to a two-dimensional
    uncorrelated random field of uniform random variables (stars), and
    two-dimensional coupled map lattices of diffusively coupled fully
    chaotic logistic maps, for coupling constant $\epsilon=0$ (squares)
    and $\epsilon=0.1$ (diamonds). The solid green line is Eq. (26) for
    $\rho=0$, the solid red line for $\rho=1$, and the solid pink line
    for $\rho=2$, (b) semi-log plot of the degree distribution of
    ILPHVG$_{8}(\rho), \rho=0,1,2$ associated to two-dimensional coupled
    map lattices of diffusively coupled fully chaotic logistic maps, for
    coupling constant $\epsilon=0.7$ (black dots), (c) the $\chi^2$
    statistic in two dimensions phase space (let time delay $\tau
    =2$), (d) scalar parameter $D$ as a function of the coupling
    constant $\epsilon$, computed from the degree distribution of
    ILPHVG$_8(\rho),\rho=0,1,2$ associated to $100\times 100$ CMLs of
    fully chaotic logistic maps.}
\end{figure}

Figs.~9(a) and 9(b) show that the degree distribution of
ILPHVG$_{8}(\rho),\rho =0,1,2$ associated with the uncoupled
($\epsilon=0$) and weakly coupled ($\epsilon=0.1$) cases is
indistinguishable from the degree distribution associated with the
$i.i.d.$ random field. Fig.~9(b) shows that the degree distribution
deviates from the theoretical result in Eq.~(26) only in the strongly
coupled case ($\epsilon=0.7$). Note that the coupled map lattices from
Eq.~(28) when $\epsilon> 0$ spatial correlations settle in and the
degree distributions of ILPHVG$_{8}(\rho)$ are statistically different
from the theoretical result in Eq.~(26), but the degree distribution of
ILPHVG$_{8}(\rho),\rho =0,1,2$ associated with the $i.i.d.$ random
field, the uncoupled case ($\epsilon=0$), and the weakly coupled case
($\epsilon=0.1$) are well approximated by Eq.~(26) in each case. There
are deviations for $k>k_{0}$ ($k_{0}=19$ for $\rho=0$, $k_{0}=36$ for
$\rho=1$, and $k_{0}=51$ for $\rho=2$) but they are caused by finite
size effects (see Fig.~7). To quantify potential deviations of the
uncoupled ($\epsilon=0$) and weakly coupled ($\epsilon=0.1$) cases from
Eq.~(26), we compute $\chi^{2}$
\begin{equation}\label{eq1}
\begin{array}{l}
\chi^{2}=N^{2}\sum\limits_{k}\frac{[P_{\rm num}(k)-P_{\rm theo}(k)]^{2}}{P_{\rm theo}(k)}
\end{array}
\end{equation}
where $P_{\rm num}(k)$ is the degree distribution of the numerical
result and $P_{\rm theo}(k)$ the theoretical result from Eq.~(26). Here
we consider 30 realizations of the $i.i.d.$ random field, the uncoupled
map lattices ($\epsilon=0$), and the weakly coupled map lattices
($\epsilon=0.1$), and in each case we use $8\leq k \leq 44$ for
$\rho=0$, $16 \leq k \leq 77$ for $\rho=1$, and $24\leq k \leq 99$ for
$\rho=2$ to compute the $\chi^2$ statistic that measures the deviation
between the empirical degree distribution and the theoretical
result. Fig.~9(c) shows the calculated results in a two-dimensional
phase space with a time delay $\tau =2$.  Note that there are clear
distinctions between the uncorrelated $i.i.d.$ random field, the
uncoupled map lattices ($\epsilon=0$), and the weakly coupled map
lattices ($\epsilon=0.1$ for $\rho=0$ and $\rho=1$), but when $\rho=2$
the distinction is no longer clear. We thus select an appropriate parameter
$\rho$ and use the degree distribution of ILPHVG$(\rho)$ to distinguish
noise from chaos.

Note that when we increase the coupling constant $\epsilon$ the
spatiotemporal dynamics of the coupled map lattice shows a rich phase
diagram. Using the degree distribution of ILPHVG$_{8}(\rho)$, we show
this rich spatiotemporal dynamics process. For each $\epsilon$ we
compute the degree distribution of the associated ILPHVG$_{8}(\rho)$. We
then compute the distance $D$ between the degree distribution at
$\epsilon$ and the corresponding result for $\epsilon=0$ in Eq.~(26),
\begin{equation}\label{eq1}
\begin{array}{l}
D=\sum\limits_{k}|P_{\rho}(k)-\frac{1}{8\rho+9}
(\frac{8\rho+8}{8\rho+9})^{k-8(\rho+1)}|,
\end{array}
\end{equation}

where $P_{\rho}(k)$ is the degree distribution of ILPHVG$_{n}(\rho)$,
and $D$ is a scalar order parameter that describes the spatial
configuration of the CML. Figure~9(d) shows that when $\rho=0,1,2$ the
evolutions of $D$ with $\epsilon$ changes from 0 to 1, indicating sharp
changes in the different phases---fully developed turbulence with weak
spatial correlations (I), periodic structure (II), spatially coherent
structure (III), and mixed structure (IV)---between periodic and
spatially-coherent structures \cite{33}. Thus the degree distribution of
the ILPHVG$_{8}(\rho)$ can capture the rich spatial structure.

\section{Discussions}

\noindent
We have introduced a directed limited penetrable horizontal visibility
graph DLPHVG$(\rho)$ and an image limited penetrable horizontal
visibility graph (ILPHVG$_{n}(\rho)$), both inspired by the limited
penetrable horizontal visibility graph LPHVG$(\rho)$ \cite{34}. These
two algorithms are expansions of the limited penetrable horizontal
visibility algorithm. We first derive theoretical results on the
topological properties of LPHVG$(\rho)$, including degree distribution
$P(k)$, mean degree $\langle k\rangle$, the relation between the datum
height $x$ and the mean degree $\langle K(x)\rangle$ of the nodes
associated to data with a height equal to $x$, the normalized mean
distance $\langle d\rangle$, the local clustering coefficient
distribution $P(C_{\rm min})$ and $P_{\rm max}$, and the probability of
long distance visibility $P_{\rho}(n)$. We then deduce the in- and
out-degree distributions $P_{\rm in}(k)$ and $P_{\rm out}(k)$ of
DLPHVG$(\rho)$, and the degree distribution of ILPHVG$_{n}(\rho)$. We
perform several numerical simulations to check the accuracy of our
analytical results. We then present applications of the directed limited
penetrable horizontal visibility graph and the image limited penetrable
horizontal visibility graph, including measuring the irreversibility of
a real-value time series and discriminating between noise and chaos, and
empirical results testify to the efficiency of our methods.

Our theoretical results on topological properties are an extension of
previous findings \cite{22,32,33,34}. In the structure of the limited
penetrable horizontal visibility graph family, the limited penetrable
parameter $\rho$ is a important and affects the structure of the
associated graphs. Under certain parameter $\rho$ values, the exact
results of the associated graphs reveals the essential characteristics
of the system, e.g., when $\rho=0$ and $\rho=1$, using the degree
distribution of ILPHVG$_{8}(\rho)$ we can distinguish between
uncorrelated and weakly coupled systems, but when $\rho=2$ the
distinction is no longer clear [see Fig.~9 (c)].  Open problem for
future research include how to use real data in selecting an optimal
limited penetrable parameter $\rho$, and how to further apply the
limited penetrable horizontal visibility graph family.

\section{Acknowledgments}

\noindent
The Research was supported by the following foundations: The National
Natural Science Foundation of China (71503132, 71690242, 91546118,
11731014, 71403105, 61403171),Qing Lan Project of Jiangsu Province
(2017), University Natural Science Foundation of Jiangsu Province
(14KJA110001), Jiangsu Center for Collaborative Innovation in
Geographical Information Resource Development and Application, CNPq, CAPES, FACEPE and UPE.

\end{document}